\documentclass[%
 aip,
 amsmath,amssymb,
 reprint,%
]{revtex4-1}

\usepackage{graphicx}
\usepackage[colorlinks=true, linkcolor=blue, citecolor=blue, urlcolor=blue]{hyperref}
\usepackage{hyperref}
\usepackage{dcolumn}
\usepackage{bm}
\usepackage{color}
\usepackage[utf8]{inputenc}
\usepackage[T1]{fontenc}
\usepackage{mathptmx}
\usepackage{etoolbox}
\usepackage{comment}
\usepackage[english]{babel}
\usepackage{amsmath}
\usepackage{threeparttable}
\usepackage{tabularx}
\usepackage[left=1.5cm, right=1.5cm, top=1.785cm, bottom=2.0cm]{geometry}

\usepackage{titlesec}

\usepackage{amssymb}
\usepackage{tikz}
\usetikzlibrary{positioning, fit, arrows.meta, calc, backgrounds}

\titlespacing*{\section}{0pt}{2ex plus 1ex minus .2ex}{1.5ex plus .2ex minus .2ex}
\titlespacing*{\subsection}{0pt}{1.5ex plus 1ex minus .2ex}{1ex plus .2ex minus .2ex}
\titlespacing*{\subsubsection}{0pt}{1.5ex plus 1ex minus .2ex}{1ex plus .2ex minus .2ex}

\makeatletter
\def\@email#1#2{%
 \endgroup
 \patchcmd{\titleblock@produce}
  {\frontmatter@RRAPformat}
  {\frontmatter@RRAPformat{\produce@RRAP{*#1\href{mailto:#2}{#2}}}\frontmatter@RRAPformat}
  {}{}
}%
\makeatother
\begin{document}

\preprint{AIP/123-QED}

\newcommand{\mytitle}{Deep Generative Markov State Models with Experimental Restraints}
\title{\mytitle}

\author{Robert M. Raddi}\noaffiliation
\author{Vincent A. Voelz\textsuperscript{*}}\noaffiliation
\affiliation{
Department of Chemistry, Temple University, Philadelphia, PA 19122, USA. \\
*Corresponding author: vvoelz@temple.edu
}%
\date{\today}

\begin{abstract}
A central challenge in molecular modeling is reconciling molecular simulations with experimental observables, as force-field inaccuracies can distort both equilibrium populations and long-timescale kinetics. While time-dependent experimental restraints can, in principle, improve the accuracy of simulated kinetics, readily accessible time-dependent structural observables remain limited. As a result, most experimental data are time-averaged, providing information about equilibrium ensembles but not directly about dynamics. Although ensemble refinement methods can improve thermodynamic accuracy, incorporating time-averaged experimental observables into kinetic models remains difficult.
Here, we introduce BICePs-reweighted Reversible DeepMSMs, a framework that combines Bayesian Inference of Conformational Populations (BICePs), variational learning of Markov processes, and maximum entropy (MaxEnt)/maximum caliber (MaxCal) principles to infer experimentally consistent thermodynamics and the minimally perturbed kinetics that accompany them.
At the core of the framework is a reversible DeepMSM prior, obtained by applying an orthogonal transformation post-hoc to a pre-trained dynamics model (e.g., VAMPnet). The construction preserves the eigenspectrum exactly while returning a valid transition matrix that is nonnegative, row-stochastic, and reversible.
The reweighted model simultaneously refines stationary populations, transition dynamics, and state-conditioned configurational landing densities within a unified framework.
We validate the approach using a quadruple-well toy system and subsequently apply it to alanine dipeptide using backbone dihedral angles and J-coupling constants as experimental observables. In both systems, perturbing the experimental observables induces predictable changes in the equilibrium ensemble, which are accompanied by corresponding changes in the inferred kinetics. The resulting relaxation timescales and dynamical modes closely agree with analytical and MaxCal reference models. Furthermore, a diffusion-based generative model of alanine dipeptide trained on the MaxEnt landing densities produces physically realistic molecular trajectories that reproduce both thermodynamics and kinetics while satisfying the imposed experimental restraints.
\end{abstract}

\maketitle


\section{Introduction}

\subsection*{The Problem}
A central challenge in computational biophysics is to model accurate biomolecular conformational dynamics that aligns with both physics-based simulations and experimental data. Despite molecular dynamics simulations greatly improving over the years, they remain limited by force field accuracy, leading to systematic errors that propagate. Even force fields trained on quantum chemical data often fail to reproduce ensemble-averaged observables without post hoc refinement\cite{boothroyd2023development}. Bridging this gap requires methods that correct simulation outputs to better agree with experimental observables.

\subsection*{Current Approaches and Limitations to Modeling Molecular Kinetics}
Markov State Models (MSMs) are widely used to model the kinetics of biomolecular systems by decomposing high-dimensional molecular simulations into networks of metastable states\cite{voelz2010molecular,bowman2026introduction}. Their utility lies in offering both kinetic and structural resolution, but constructing accurate MSMs often involves complex, multi-step workflows—including feature selection, dimensionality reduction, clustering, and lag time estimation—that can introduce systematic errors and bias.

To address these challenges, deep learning-based approaches such as the Variational Approach for Markov Processes (VAMP)\cite{wu2020variational} and its extension, VAMPnets,\cite{mardt2018vampnets} have been developed to automate MSM construction. VAMPnets unify feature extraction, dimensionality reduction, and state decomposition into a single end-to-end framework, trained by variationally maximizing the VAMP2 score (e.g.) to identify slow collective modes. VAMPnets have also been applied in enhanced sampling,\cite{kleiman2023active} latent space simulators,\cite{sidky2020molecular, jones2023molecular} and other workflows aimed at efficient modeling of slow dynamics and generative simulation. However, like traditional MSMs, VAMPnets remain dependent on the accuracy of the underlying force field.

\subsection*{Post-Processing Approaches for Reconciling Simulations with Experiments}
Various methods have been developed to address the challenge of reconciling simulation predictions with ensemble-averaged experimental observables, employing either maximum entropy (MaxEnt)\cite{roux2013statistical,cavalli2013molecular}, Bayesian inference\cite{perez2015accelerating,fisher2010modeling,rieping2005inferential}, or a combination of the two \cite{orioli2020learn,bonomi2016metainference,bonomi2017principles, boomsma2014combining,beauchamp2014bayesian,brookes2016experimental,raddi2025model, cesari2018using,hummer2015bayesian}. Although some of these methods apply restraints during simulation, we are particularly interested in post-processing techniques that reweight conformational ensembles derived from molecular simulations. Such reweighting methods aim to correct thermodynamic and kinetic properties while accounting for uncertainties, independently of the simulation protocol itself.

Examples of this approach include BioEn\cite{hummer2015bayesian, kofinger2019inferring} and BME,\cite{pesce2021refining,bottaro2020integrating}, which apply MaxEnt principles while constraining a $\chi^2$ metric to characterize simulation-experiment discrepancies. Augmented Markov models (AMMs)\cite{olsson2017combining} extend this idea by incorporating experimental observables into the MSM construction process. However, these methods often demand extensive hyperparameter tuning, and knowledge of experimental and forward model uncertainties \textit{a priori} which are typically unknown. Moreover, we desire reweighting strategies that not only refine thermodynamic populations but also enable improved kinetic modeling of the conformational landscape in a \textit{post hoc} fashion.

\subsection*{The BICePs Framework}
The Bayesian Inference of Conformational Populations (BICePs) algorithm\cite{raddi2025model,raddi2024thesis} offers a unique MaxEnt-based post-processing solution that circumvents the need for predefined experimental uncertainties by inferring them directly from the data.\cite{Voelz:2014fga,voelz2021reconciling,liang2018elucidating,Ge:2018bdb,hurley2021,raddi2023biceps,raddi2024thesis}   By combining high-quality theory- or simulation-based conformational ensembles with sparse experimental observables as a post-processing step, BICePs allows sampling over the posterior distribution of conformational states while accounting for both random and systematic errors, with improved likelihoods to deal with outliers \cite{raddi2023biceps}. This versatility has enabled applications in force field optimization \cite{raddi2026automated}, forward model optimization \cite{Raddi2024FMO}, and model selection \cite{raddi2025model}. For example, we have used BICePs to evaluate and rank force fields for the mini-protein chignolin by their agreement with experimental NMR measurements. Additionally, we have shown that BICePs reweighting is able to fine-tune the folding landscapes of non-natural cyclic and linear peptides, vastly improving predictions of folding stability\cite{NRV2024}.  However, in all of our previous work, BICePs has primarily been used to correct thermodynamic properties, ignoring changes in kinetics.

\subsection*{Deep Reversible Generative Markov State Models Informed by Experimental Data}
In this work, we construct deep Markov models\cite{mardt2018vampnets,wu2018deep,mardt2021progress} whose thermodynamics and kinetics are learned by reconciling prior simulation data with information provided by experimental observations. More generally, our framework is compatible with any maximum-entropy (MaxEnt) or Bayesian reweighting procedure that produces conformational ensembles consistent with both simulation and experiment.\cite{gilardoni2025mdrefine,kofinger2019inferring,pesce2021refining,bottaro2020integrating,olsson2017combining} In the present work, we use Bayesian Inference of Conformational Populations (BICePs) as the reweighting method because it provides a MaxEnt post-processing framework that avoids the need to specify experimental uncertainties a priori, instead treating them as nuisance parameters to be inferred from the data.  We demonstrate that our BICePs-reweighted Deep Reversible MSM framework can uncover accurate dynamics and account for systematic force field error using time-averaged experimental restraints. We first validate our approach using a quadruple-well toy system, and then apply it to a more realistic alanine dipeptide system.

\section{Theory}

To integrate experimental restraints into the modeling of molecular kinetics, we adopted and modified an end-to-end trainable deep learning framework\cite{mardt2018vampnets,wu2018deep} that learns how molecular systems evolve over time using both simulation data and structural information derived from experiments. As illustrated in Figure \ref{fig:flowchart}, the approach takes as input pairs of molecular configurations $x_{t}, x_{t+\tau}$ separated by a fixed lag time $\tau$ and processes them through a neural network, referred to as the encoder. The encoder assigns each configuration a soft membership $\chi$ across a set of metastable states designed to capture the slow dynamical modes of the system.

From these ``fuzzy'' state definitions, we construct two objects that our framework couples. The first is a prior transition model. This is constructed from a Koopman operator estimate obtained from the membership covariances and time-lagged covariances.  The transition model conserves probability and already carries the correct stationary vector; its only defect as a transition matrix is that its entries may be negative. We remove this defect \emph{post hoc}, by an orthogonal transformation in whitened coordinates that leaves the variational (VAC) spectrum exactly invariant while placing the model on a probability simplex, yielding a nonnegative, row-stochastic, reversible transition matrix $T^{(0)}$ and its symmetric equilibrium flux $F^{(0)}=\mathbf{D}_{\pi_0}T^{(0)}$ (see below). The second object is our framework is the reweighted stationary distribution $\bm\pi$, which is inferred from experimental restraints using the BICePs approach.   With the  memberships $\chi$ held fixed,  BICePs infers state populations $\bm\pi$ consistent with the experimental observables, together with their posterior covariance, marginalizing over all nuisance uncertainty parameters.

Next, to construct a Reweighted Reversible MSM that predicts \textit{kinetics} consistent with the reweighted populations $\bm\pi$, the prior transition model is transformed according to the Maximum Caliber (MaxCal) principle\cite{dixit2015inferring,ghosh2020maximum} , which imposes $\bm\pi$ on $F^{(0)}$ as a congruence transformation of the flux. Any target distribution---and any perturbation to prior populations from experimental data---can thus be applied to a single trained network without retraining. Because the congruence acts on the flux rather than on the spectrum, the implied timescales respond to the reweighted populations rather than being held fixed by construction. The resulting model is a valid, $\bm\pi$-reversible transition matrix whose state-conditioned landing densities follow in closed form, and which may be used either to reweight and resample existing simulation data, or to train a generative decoder producing entirely new configurations consistent with the experimentally restrained dynamics. This framework provides a principled way to generate dynamical models that are informed by both physics-based simulations and experimental measurements.

\begin{figure}[!th]
\centering
\begin{tikzpicture}[
    font=\normalsize,
    b/.style   ={draw, rounded corners=2pt, semithick, inner sep=3.5pt,
                 align=center, text width=7.55cm},
    b1/.style   ={draw, rounded corners=2pt, semithick, inner sep=3.5pt,
                 align=center, text width=7.8cm},
    h/.style   ={b, text width=3.5cm},
    data/.style={h, fill=gray!12,    draw=gray!70!black},
    model/.style={b1, fill=blue!8,    draw=blue!60!black},
    prior/.style={h, fill=cyan!10,   draw=cyan!55!black, minimum height=2.6cm},
    exp/.style ={h, fill=green!10,   draw=green!55!black, minimum height=2.6cm},
    core/.style={b, fill=purple!10,  draw=purple!60!black},
    gen/.style ={b, fill=orange!14,  draw=orange!75!black},
    a/.style   ={-{Latex[length=2.2mm,width=1.7mm]}, line width=0.9pt,
                 shorten >=1.5pt, shorten <=1.5pt},
    l/.style   ={fill=white, inner sep=1.2pt, font=\normalsize}
]
\node[data] (traj) {\textbf{MD trajectories}\\[2pt] $\{(x_t,x_{t+\tau})\}$};
\node[data, right=0.5cm of traj] (exp)
  {\textbf{Experimental data}\\[2pt] $D$};
\node[model, below=1.0cm of $(traj)!0.5!(exp)$] (feat) {%
  \textbf{Dynamics model}\ \ {VAMP, MEMnets, DeepMSM, \textbf{\dots}}\\[3pt]
  fuzzy memberships $\chi_i(x)\ge 0$, \ $\sum_i\chi_i(x)=1$\\[3pt]
  };
\node[prior, below=1.0cm of feat, xshift=-2.02cm] (pri) {%
  \textbf{RevDeepMSM prior} \\[3pt]
  (exact VAC spectrum)\\[3pt]
  $F^{(0)}\!=\!\mathbf{D}_{\pi_0}T^{(0)}$\\[3pt]
  nonnegative symmetric flux\\[3pt]
  };
\node[exp, below=1.0cm of feat, xshift=2.02cm] (bic) {%
  \textbf{BICePs}\\[3pt]
  reweighted populations\\[4pt]
  $\bm\pi=\!\displaystyle\int\! p(\mathbf{X},\sigma\mid D)\,d\sigma$\\[4pt]
  posterior covariance $\bm\Sigma_\pi$
  };
\begin{scope}[on background layer]
  \node[draw=gray!45, dashed, rounded corners=4pt, fill=gray!4,
        inner xsep=4pt, inner ysep=4pt, fit=(pri)(bic)] (band) {};
\end{scope}
\node[core, below=1.0cm of band] (msm) {%
  \textbf{Reweighted Reversible DeepMSM}\\[4pt]
  $F^{(0)} \; {\Large \xrightarrow[ \mbox{\normalsize\text{target }$\pi$} ]{ \mbox{\normalsize\text{MaxCal}} } } \;  T$ \\[4pt]
  $\displaystyle q_i(x_{t+\tau})=\mu(x_{t+\tau})\sum_j\frac{T_{ij}\,\chi_j(x_{t+\tau})}{\pi_j}$\\[4pt]
  };
\node[gen, below=0.8cm of msm] (gen) {%
  \textbf{Generative decoder}\ \ {\normalsize(diffusion / flow)}\\[3pt]
  $\tilde{x}_{t+\tau}\sim p_\theta(x\mid X_i)$};
\draw[a] (traj) -- (traj |- feat.north);
\draw[a] (exp.east) -- ++(0.23,0) |- (bic.east);
\draw[a] (feat.south -| pri) --
  node[l,pos=0.35]{$\mathbf{C}_{00},\mathbf{C}_{01}$} ([yshift=1pt]pri.north);
\draw[a] (feat.south -| bic) -- node[l,pos=0.35]{$\chi$} ([yshift=1pt]bic.north);
\draw[a] ($(pri.north)+(1.0,0)$) -- ++(0,0.46)
  -- node[above, inner sep=1pt]{$\bm\pi_0$}
  ($(bic.north)+(-1.0,0.46)$) -- ($(bic.north)+(-1.0,0)$);
\draw[a] (band.south -| pri) -- node[l,pos=0.35]{$F^{(0)}$} ([xshift=-2.02cm]msm.north);
\draw[a] (band.south -| bic) -- node[l,pos=0.35]{$\bm\pi\pm\bm\sigma_\pi$}
        ([xshift=2.02cm]msm.north);
\draw[a] (msm) -- node[l,pos=0.35]{$q_i$} (gen);
\end{tikzpicture}
\caption{\small Schematic overview of the framework for learning a reweighted, reversible deep MSM consistent with molecular simulation data and experimental restraints. Trajectory pairs $(x_t,x_{t+\tau})$ train a dynamics model whose fuzzy state memberships $\chi$ define the instantaneous and time-lagged covariances. An orthogonal rotation in whitened coordinates converts the Koopman estimate $K^{(0)}=\mathbf{C}_{00}^{-1}\mathbf{C}_{01}$ into a symmetric, nonnegative flux $F^{(0)}$ carrying the VAC spectrum exactly. BICePs reweights the state populations given experimental data $D$, and Maximum Caliber imposes $\bm\pi$ as a congruence on the flux, $F=\mathbf{D}_\alpha F^{(0)}\mathbf{D}_\alpha$. The resulting transition matrix is nonnegative, row-stochastic and $\bm\pi$-reversible by construction, and its landing densities $q_i$ follow in closed form. These landing densities may be used to resample training configurations or to train a diffusion decoder generating novel configurations $\tilde{x}_{t+\tau}\sim p_\theta(x\mid X_i)$ consistent with the experimental restraints.}
\label{fig:flowchart}
\end{figure}

\subsection{Reweighted Reversible DeepMSM}
\label{sec:deepMSM}

\subsubsection*{Building an Initial Transition Model from Trajectory Data.}
Given time-lagged configurations $x_{t}$ and $x_{t+\tau}$, we encode the mapping of molecular coordinates to fuzzy Markov states. The resultant dimensionality reduction $\chi(x)\in\mathbb{R}^m$ represents the probability of a configuration $x$ to be in a metastable state $X_i$
\begin{equation}
  \begin{aligned}
    \chi_{i}(x_{t}) =& \; p(x_{t} \in X_{i} \mid x_{t}) \\
    \chi_{i}(x_{t+\tau}) =& \; p(x_{t+\tau} \in X_{i} \mid x_{t+\tau})
  \end{aligned}
  \label{eq:state_membership}
\end{equation}
These membership probabilities are nonnegative ($\chi_{i}(x) \ge 0 \, \forall x$) and sum to one ($\sum_{i} \chi_{i}(x) = 1 \, \forall x$), which is achieved by constructing the network with the output layer consisting of Softmax output nodes. Similar to VAMPnets\cite{mardt2018vampnets}, the $\chi$ network parameters are learned by passing batches of training data through the network while maximizing a VAMP score. We emphasize, however, that the memberships may come from any pre-trained dynamics model, and that everything below treats $\chi$ as fixed.

\paragraph{Constructing a reweighted Koopman matrix $K^{(0)}$. } Using the membership functions $\chi$ , we next construct a Koopman matrix from the trajectory data $x_t$, which requires estimation of the instantaneous and time-lagged covariances $\tilde{\mathbf{C}}_{00}=\mathbb{E}_t[\chi(x_t)\chi(x_t)^{\top}]$ and $\tilde{\mathbf{C}}_{01}=\mathbb{E}_t[\chi(x_t)\chi(x_{t+\tau})^{\top}]$, where $\tau$ is the lag time of the model. To correct for off-equilibrium sampling, we use the Koopman reweighting approach of Wu \textit{et al.}\cite{wu2017variational} and Mardt \textit{et al.}\cite{mardt2020deep}, in which each transition pair $(x_t,x_{t+\tau})$ carries a weight $w_t=\chi(x_t)^{\top}u$, with $u$ fixed by requiring the reweighted ensemble to be stationary,
\begin{equation}
  \sum_t w_t \left[ \chi(x_{t+\tau}) - \chi(x_t) \right] = 0 .
  \label{eq:stationarity}
\end{equation}
In terms of the instantaneous and time-lagged covariances $\tilde{\mathbf{C}}_{00}$ and $\tilde{\mathbf{C}}_{01}$, Eq. \ref{eq:stationarity} reads $\tilde{\mathbf{C}}_{01}^{\top}u=\tilde{\mathbf{C}}_{00}u$ and is solved by $u=\tilde{\mathbf{C}}_{00}^{-1}\hat{\bm\pi}$, with $\hat{\bm\pi}$ the left eigenvector of $\tilde{\mathbf{C}}_{00}^{-1}\tilde{\mathbf{C}}_{01}$ at eigenvalue one. The weights must also be nonnegative, since only nonnegative weights keep $\mathbf{C}_{00}$ positive definite, as the similarity transform step below requires. We therefore solve the least-squares form of Eq. \ref{eq:stationarity} given in Ref. \cite{wu2017variational} with the nonnegativity constraint added, a convex nonnegative least-squares problem in $m$ variables
\begin{equation}
  u = \arg\min_{u \ge 0} \left\| \tilde{\mathbf{C}}_{01}^{\top}u - \tilde{\mathbf{C}}_{00}u \right\|^{2},
  \label{eq:nnls}
\end{equation}
with $\mathbf{1}^{\top}\tilde{\mathbf{C}}_{00}u = 1$, which for memberships summing to one is the normalization $\sum_t w_t = 1$ of Ref. \cite{wu2017variational}. The normalization only sets the scale of $u$ and rules out the trivial solution, since every expectation below is a ratio, $\mathbb{E}^{w}[\,\cdot\,]=\sum_t w_t(\cdot)/\sum_t w_t$. Whenever the unconstrained solution is already nonnegative, the two agree exactly, so Eq. \ref{eq:nnls} enforces a constraint of the model rather than modifying the estimator, and $w_t \ge 0$ holds by construction with no truncation. For trajectory data sampled at equilibrium, $w$ is constant in the large-data limit and the unweighted estimator is recovered. We form time-reversal-symmetrized covariances by averaging each estimator over the forward and reversed transition pairs,
\begin{equation}
  \begin{aligned}
  \mathbf{C}_{00} &= \tfrac{1}{2}\,\mathbb{E}^w\!\left[ \chi(x_t)\chi(x_t)^{\top} + \chi(x_{t+\tau})\chi(x_{t+\tau})^{\top}\right],\\
  \mathbf{C}_{01} &= \tfrac{1}{2}\,\mathbb{E}^w\!\left[ \chi(x_t)\chi(x_{t+\tau})^{\top} + \chi(x_{t+\tau})\chi(x_t)^{\top}\right],
  \end{aligned}
\label{eq:covariances}
\end{equation}
so that $\mathbf{C}_{00}=\mathbf{C}_{00}^{\top}$ and $\mathbf{C}_{01}=\mathbf{C}_{01}^{\top}$ by construction. From these quantities, the Koopman matrix is estimated as
\begin{equation}
K^{(0)}=\mathbf{C}_{00}^{-1}\mathbf{C}_{01}.
\label{eq:koopman}
\end{equation}
Because the memberships form a partition of unity, $\mathbf{C}_{00}\mathbf{1}=\mathbf{C}_{01}\mathbf{1}=\bm\pi_0$.  Furthermore, the following properties hold:  $K^{(0)}\mathbf{1}=\mathbf{C}_{00}^{-1}\bm\pi_0=\mathbf{1}$ and $(K^{(0)})^{\!\top}\bm\pi_0=\bm\pi_0$.  Therefore, the Koopman estimate conserves probability mass and already has $\bm\pi_0$ as its stationary vector. Its \emph{only} defect as a transition matrix is that its entries may be negative—this is a deficiency that motivated the DeepMSM framework of Wu \textit{et al.}\cite{wu2018deep}. To achieve a valid transition matrix, we must remove the negative entries without altering the eigenspectrum, which a \textit{similarity} transformation of $K^{(0)}$ followed by orthogonal rotation achieves.

\subsubsection{Similarity transform of $K^{(0)}$ in a whitened basis.}
Although the Koopman matrix $K^{(0)}$ is self-adjoint in the $\mathbf{C}_{00}$ inner product by Eq. \ref{eq:covariances}, this metric is tilted by the overlap encoded in the off-diagonal structure of $\mathbf{C}_{00}$.  Whitening removes this tilt by rescaling the features as $\tilde\chi=\mathbf C_{00}^{-1/2}\chi$, so that their instantaneous covariance becomes the identity. In this basis, the propagator is
\begin{equation}
    \mathbf{S}_0 =\mathbf{C}_{00}^{-1/2}\,\mathbf{C}_{01}\,\mathbf{C}_{00}^{-1/2}   =\mathbf{C}_{00}^{1/2}\,K^{(0)}\,\mathbf{C}_{00}^{-1/2},
\label{eq:S0}
\end{equation}
which is a similarity transform of $K^{(0)}$ and therefore preserves its VAC eigenspectrum, and hence its implied timescales, exactly. Because $\mathbf{C}_{01}$ is symmetric, $\mathbf{S}_0$ is symmetric in the ordinary Euclidean inner product, with real eigenvalues and orthonormal eigenvectors.  The symmetrization in Eq. \ref{eq:covariances} ensures that $\mathbf{S}_0$ has a real spectrum. Its eigenvalues coincide with those obtained from the variational approach for conformational dynamics (VAC), and we therefore refer to them as the VAC spectrum throughout. In contrast, VAMP is formulated in terms of singular values: without an assumption of reversibility, the Koopman operator is not self-adjoint, and its eigenvalues do not satisfy an analogous variational bound.

This makes orthogonal rotations the natural spectrum-preserving transforms in the whitened basis, while preserving symmetry and reversibility. Additionally, the dominant eigenvector is the unit vector $c=\mathbf{C}_{00}^{1/2}\mathbf{1}$, since $\mathbf{S}_0c=c$ and $\lVert c\rVert^2=\mathbf{1}^{\top}\mathbf{C}_{00}\mathbf{1}=\mathbf{1}^{\top}\bm\pi_0=1$.

\subsubsection{An orthogonal rotation repairs transition matrix validity at no spectral cost: the Reversible DeepMSM prior.}
In the whitened basis, orthogonal transformations preserve both the symmetry and the spectrum of $\mathbf{S}_0$. Thus, for any $R^{\top}R=\mathbf{I}$, $\mathbf{S}=R\,\mathbf{S}_0R^{\top}$ is symmetric and has exactly the same eigenvalues as $\mathbf{S}_0$.  We choose $R$ such that
\begin{equation}
R\,c=\sqrt{\bm\pi_0},
\label{eq:R}
\end{equation}
which is always possible because $c$ and $\sqrt{\bm\pi_0}$ are both unit vectors, and therefore $\mathbf{S}\sqrt{\bm\pi_0}=\sqrt{\bm\pi_0}$. Undoing the whitening by transforming back to transition-matrix coordinates, with $\mathbf{D}_{\pi_0} = \operatorname{diag}(\pi_{0})$,
\begin{equation}
T^{(0)}=\mathbf{D}_{\pi_0}^{-1/2}\,\mathbf{S}\,\mathbf{D}_{\pi_0}^{1/2}
\end{equation}
gives $T^{(0)} \mathbf{1} = \mathbf{1}$, while
\begin{equation}
  F^{(0)}=\mathbf{D}_{\pi_0}T^{(0)} = \mathbf D_{\pi_0}^{1/2}\mathbf S\mathbf D_{\pi_0}^{1/2}
\label{eq:F0}
\end{equation}
is symmetric, so $T^{(0)}$ is reversible with stationary distribution $\bm\pi_{0}$.  At this stage, row normalization and reversibility are guaranteed, while entry-wise nonnegativity is the only remaining condition for $T^{(0)}$ to be a valid transition matrix. Importantly, the condition $Rc = \sqrt{\bm\pi_{0}}$ leaves $R$ non-unique: any transformation $R^\prime=RQ$, where $Q$ is orthogonal and satisfies $Qc=c$, preserves the same mapping. For $m$ states, this residual freedom forms an $O(m-1)$ family with $(m-1)(m-2)/2$ free parameters, which can be optimized to minimize violations of nonnegativity without changing the stationary distribution, reversibility, or eigenspectrum.  The resulting $T^{(0)}$ is therefore a valid reversible transition matrix with the VAC spectrum preserved exactly.   Equivalently, its flux $F^{(0)}=\mathbf{D}_{\pi_0}T^{(0)}$ is a symmetric joint distribution over state pairs satisfying $F^{(0)}\mathbf{1}=\bm\pi_0$ and $\mathbf{1}^{\top}F^{(0)}\mathbf{1}=1$.

In the crisp limit, i.e. when $\chi_i(x)$ are binary state indicator functions, $\mathbf{C}_{00}\to\mathbf{D}_{\pi_0}$, so $c=\sqrt{\bm\pi_{0}}$, $R=\mathbf I$, and the construction reduces to $T^{(0)}=K^{(0)}$. Away from this limit, the orthogonal transformation uses the residual freedom of the whitened representation to obtain a physically admissible transition matrix without altering the dynamical spectrum.  These eigenvalues solve the generalized eigenvalue problem $\mathbf{C}_{01}v=\lambda\,\mathbf{C}_{00}v$ and are variationally optimal within $\mathrm{span}\{\chi\}$. We have therefore constructed a reversible DeepMSM, \textit{prior to reweighting}, that is normalized, reversible, and nonnegative while retaining the VAC eigenspectrum exactly.  Unlike imposing these constraints during training, this post-hoc construction preserves the learned VAC eigenspectrum by design (Sec. \ref{sec:mardt-comparison}).

\subsubsection{The BICePs algorithm provides reweighted populations}
The BICePs algorithm \cite{raddi2025model} employs a Bayesian framework to model the joint posterior distribution $p(\mathbf{X}, \sigma \mid D)$ of a set of $N_{r}$ replicas of conformational states $\mathbf{X} = \{X_r \text{ for } r = 1, \ldots, N_r\}$ and a set of uncertainty parameters $\sigma = \{\sigma_j \text{ for } j = 1, \ldots, N_d\}$, given experimental data $D$, which consists of $N_d$ ensemble-averaged measurements. The uncertainty parameters quantify finite-sampling error in the simulations as well as uncertainty in the experimental measurements and the forward model error used to predict observables from structural configurations.  Because the BICePs algorithm has been described extensively in previous work;\cite{raddi2025model,Raddi2024FMO,NRV2024} we provide a brief overview below.

According to Bayes’ theorem, the posterior distribution is given by
\begin{equation}
p(\mathbf{X},\sigma | D) \propto p(D | \mathbf{X},\sigma)p(\mathbf{X}) p(\sigma)
\end{equation}
where $p(D \mid \mathbf{X}, \sigma)$ is the likelihood of observing the data given values of $\mathbf{X}$ and $\sigma$, $p(\mathbf{X}) = \prod_r p(X_r)$ encodes prior knowledge from the Koopman model built from soft state membership probabilities for all $m$ states $\{\chi_{i}\}_{i=1}^{m}$ (Eq. \ref{eq:state_membership}) given all molecular simulation data, and $p(\sigma) = \prod_j p(\sigma_j) \sim \sigma_j^{-1}$ are noninformative Jeffreys priors.

The posterior $p(\mathbf{X}, \sigma \mid D)$ is sampled using Markov Chain Monte Carlo (MCMC), with the likelihood function serving to reweight the sampled configurations $X_r$ to match ensemble-averaged experimental measurements. This is done using a forward model that maps conformational states to predicted observables. In the limit of large numbers of replicas, BICePs approaches the maximum entropy solution over the predicted ensemble, making it a MaxEnt reweighting algorithm.\cite{pitera2012use,cavalli2013molecular,cesari2018using,roux2013statistical,hummer2015bayesian}.

\paragraph*{The forward model.}
To quantify the agreement between the experiment and our simulations, we require a \textit{forward model}, which predicts experimental observables from simulated conformations. The forward model data $g$ for state $X_{i}$ is a weighted average over all configurations $x$ using either a soft encoding or a one-hot encoding of their membership probabilities:
\begin{equation}
  g_{j}(X_{i}) = \frac{1}{\sum_{t=1}^{T-\tau} \chi_{i}(x_{t})} \sum_{t=1}^{T-\tau} \chi_{i}(x_{t}) g_{j}(x_{t}).
\end{equation}

\paragraph*{The likelihood model.}
In its most fundamental form, BICePs employs a Gaussian likelihood function with an independent uncertainty parameter $\sigma_j$ for each experimental observable $d_j$. Each $\sigma_j$ is treated as a nuisance parameter that capture both measurement error and finite-sampling noise in the forward model prediction. The Gaussian likelihood over all observables is
\begin{equation}
  p(D \mid \mathbf{X}, \mathbf{\sigma}) = \prod_{j=1}^{N_{d}} \mathcal{N}\left(g_j(\mathbf{X}) \,\middle|\, d_j, \sigma_{j}^2 \right),
  \label{eq:gaussian_likelihood}
\end{equation}
where $g_j(\mathbf{X}) = \frac{1}{N_r} \sum_{r=1}^{N_r} g_j(X_r)$ is the replica-averaged forward model prediction, and $\sigma_{j}^2 = (\sigma_{j}^B)^2 + (\sigma_{j}^{\text{SEM}})^2$ combines Bayesian prior uncertainty and finite-sampling error. The standard error of the mean is estimated from replica fluctuations as $\sigma_j^{\text{SEM}} = \sqrt{\frac{1}{N_r} \sum_{r=1}^{N_r} \left(g_j(X_r) - \langle g_j(\mathbf{X}) \rangle\right)^2}$, which decreases as $N_r^{-1/2}$ and enables more accurate modeling of ensemble averages.  More sophisticated likelihood functions, described in Ref. \cite{raddi2025model}, can be used to better model experimental outliers.

\paragraph*{Posterior state populations.}
The reweighted populations are obtained from the marginal distribution
\begin{equation}
  \mathbf{\pi} = \int p(\mathbf{X}, \sigma \mid D) d\sigma.
  \label{eq:posterior_populations}
\end{equation}
A key advantage of this approach is that it obviates the need for an empirical regularization parameter to balance the contributions of the prior model and experimental data.

\subsubsection{Maximum Caliber imposes the reweighted $\pi$ as a constraint on the flux.}
Given the physically admissible prior $T^{(0)}$, we define its equilibrium flux as $F^{(0)}=\mathbf D_{\pi_0}T^{(0)}$. An external target stationary distribution $\bm\pi$, here the BICePs posterior mean, is then imposed by finding the flux $F$ that minimally perturbs $F^{(0)}$ using the relative entropy objective
\begin{equation}
\min_{F}\;D_{\mathrm{KL}}\!\left(F\,\middle\|\,F^{(0)}\right),
\label{eq:maxcal}
\end{equation}
subject to the following constraints: (i) nonnegativity $F\ge 0$, (ii) detailed balance $F=F^{\top}$, and (iii) stationary populations $F\mathbf 1=\bm\pi$.  The solution has the symmetric scaling form
\begin{equation}
F=\mathbf D_{\alpha}F^{(0)}\mathbf D_{\alpha},
\end{equation}
where $\alpha$ is determined by $\alpha_i(F^{(0)}\alpha)_i=\pi_i$ and can be obtained by symmetric Sinkhorn iteration \cite{sinkhorn1967,dixit2015inferring,ghosh2020maximum}. The reweighted transition matrix is then
\begin{equation}
T=\mathbf D_{\pi}^{-1}F,
\label{eq:Treweighted}
\end{equation}
which satisfies $T\mathbf 1=\mathbf 1$ and detailed balance with respect to $\bm\pi$.  Unlike the orthogonal transformation used to construct $T^{(0)}$, this operation is a congruence transformation of the flux that can change the kinetic spectrum, allowing the implied timescales to respond to the reweighted populations. Writing $r_i=\pi_i/\pi^{(0)}_i$ for the population reweighting factors and $\alpha_i=\sqrt{r_i}\,a_i$, the reweighted rates obey
\begin{equation}
\frac{T_{ij}}{T^{(0)}_{ij}} = a_i a_j\left(\frac{r_j}{r_i}\right)^{1/2},
\label{eq:sqrt-scaling}
\end{equation}
where $T_{ij}$ and the symmetric factor $a_i a_j$ enforces $F\mathbf 1=\bm\pi$.  Note that Eq. \ref{eq:sqrt-scaling} implies the square-root relationship $T_{ij} \propto \sqrt{\pi_j/\pi_i}$, a result of Maximum Caliber\cite{dixit2015inferring} that independently arises from the so-called square-root approximation arising from finite-volume discretization of the Fokker--Planck operator \cite{lie2013square,donati2021markov,donati2022review}

\subsubsection{MaxEnt landing densities in closed form.}
The landing densities $q_i(x_{t+\tau}; \boldsymbol{\pi})$ model the distribution of configurations at time $t+\tau$, after originating in state $i$ at time $t$.  The landing densities are implicitly conditioned on experimental data $D$ through the inferred reweighted state populations $\boldsymbol{\pi}$, and the system occupying metastable state $X_i$ at time $t$:
\begin{equation}
  q_{i}(x_{t+\tau}; \boldsymbol{\pi}) = p(x_{t+\tau} \mid x_{t} \in X_{i}; \boldsymbol{\pi})
  \label{eq:landing_densities}
\end{equation}
These probability densities satisfy $q_i(x) \ge 0 \ \forall x$ and are normalized over observed configurations $\int q_i(x_{t+\tau}) \, dx_{t+\tau} = 1 \ \forall i$. Each $q_i$ defines a predictive distribution of the time evolution of configurations originating in state $X_i$ state, after a lag time $\tau$.

To obtain the landing densities $q_i(x_{t+\tau}; \boldsymbol{\pi})$, we must consider the continuous reweighted configurational density $\mu(x)$.  We require $\mu$ be positive, normalized, and consistent with the reweighted populations, $\mathbb{E}_\mu[\chi]=\bm\pi$. While $\mu$ is not known in advance, these requirements put strong constraints on $\mu$, enabling its determination, described below.

We use $\mu$ to develop a model for the landing densities as follows:  Because the state membership functions $\chi_i$ are conditional state probabilities, Bayes' rule gives the emission density $p(x_{t+\tau}\mid x_{t+\tau}\in X_j)=\chi_j(x_{t+\tau})\mu(x_{t+\tau})/\pi_j$. Marginalizing over the state occupied at $t+\tau$, and invoking the Markov assumption that the configuration depends on the past only through the state it occupies,
\begin{equation}
  q_i(x_{t+\tau};\bm\pi) =\sum_j p(x_{t+\tau}\mid x_{t+\tau}\in X_j)\,T_{ij},
  \label{eq:landing_marginal}
\end{equation}
where $T_{ij} = p(x_{t+\tau}\in X_j | x_{t}\in X_i)$.
Substituting the emission density yields the closed form
\begin{equation}
\begin{aligned}
q_i(x_{t+\tau};\bm\pi) &=\mu(x_{t+\tau})\sum_j\frac{T_{ij}\,\chi_j(x_{t+\tau})}{\pi_j}\\
  &=\mu(x_{t+\tau})\left[\mathbf{S}\,\chi(x_{t+\tau})\right]_i,
\end{aligned}
\label{eq:landing}
\end{equation}
where $\mathbf{S}=\mathbf{D}_\pi^{-1}F\,\mathbf{D}_\pi^{-1}$. Normalization is inherited from the moment constraint, $\int q_i\,dx=\sum_j T_{ij}\,\mathbb{E}_\mu[\chi_j]/\pi_j=\sum_j T_{ij}=1$. Since $F$ is symmetric and nonnegative, so is $\mathbf{S}$, placing the model in the reversible, nonnegative class of Mardt \textit{et al.} by construction\cite{mardt2020deep}.

Since $\mu$ is not known in advance, we obtain it through enforcing self-consistency of the model rather than imposing it separately. Given a current estimate $\mu^{(k)}$, the current estimate of Eq. \ref{eq:landing} gives the corresponding $q^{(k)}$, and the stationary density is updated to what those densities provide,
\begin{equation}
  \mu^{(k+1)}(x_{t+\tau}) \;\leftarrow\; \sum_i \pi_i\,q^{(k)}_i(x_{t+\tau};\bm\pi),
  \label{eq:mu_fixedpoint}
\end{equation}
starting from the empirical distribution of sampled configurations $\mu^{(0)}$. Convergence is reached within a few iterations, thereby resulting in $\mathbb{E}_\mu[\chi] = \bm\pi$, where the landing densities are exactly normalized.

The membership overlap relates the landing densities to the transition matrix. Taking membership moments of Eq. \ref{eq:landing},
\begin{equation}
  \int q_i(x_{t+\tau})\,\chi_j(x_{t+\tau})\,dx_{t+\tau} = \left[T\,\bm\Phi\right]_{ij},
  \label{eq:landing_blur}
\end{equation}
where $\bm\Phi=\mathbf{D}_\pi^{-1}M$ and $M_{jk}=\mathbb{E}_\mu[\chi_j\chi_k]$. The matrix $\bm\Phi$ is the same membership overlap that tilts $\mathbf{C}_{00}$ in Eq. \ref{eq:koopman}, evaluated under the reweighted density rather than the empirical one. This is a property of the state decomposition itself and is present in any fuzzy-state model. It is a valid $\bm\pi$-reversible transition matrix in its own right, satisfying $\bm\Phi\ge 0$, $\bm\Phi\mathbf{1}=\mathbf{1}$ and $\bm\pi^{\top}\bm\Phi=\bm\pi^{\top}$, and describes one emission--reencoding cycle. When the memberships are crisp, $\chi_j\chi_k=\delta_{jk}\chi_j$, so $M=\mathbf{D}_\pi$, $\bm\Phi=\mathbf{I}$, and the kinetics are reproduced exactly.

Eq. \ref{eq:landing_blur} gives $T\bm\Phi$ rather than $T$. The overlap is the entire systematic error between the transition matrix we estimate and the one recovered from a cycle of emission and reencoding, and $\bm\Phi=\mathbf{I}$ removes it. Since $\bm\Phi$ departs from the identity only through the off-diagonal moments $\mathbb{E}_\mu[\chi_j\chi_k]$, which vanish exactly when no configuration carries membership in two states at once, we report the crispness of the memberships,
\begin{equation}
\kappa \left[ \chi(x_t) \right] = \frac{ \mathbb{E}_t \! \left[ \max_{i} \chi_i(x_t) \right] - 1/m}{1-1/m},
\end{equation}
which ranges from $0$ for uniform assignments to $1$ for hard assignments. Crispness is a measured property of the trained model rather than an assumption, and $\kappa\to 1$ drives $\bm\Phi\to\mathbf{I}$. Metastability helps, because a slow process is one that stays inside a basin for a long time and crosses between basins rarely, so a network trained to capture it assigns nearly all of the membership of each configuration to a single basin. In the DeepGenMSM construction of Wu \textit{et al.}\cite{wu2018deep} the transition matrix is defined through this same cycle, so $\bm\Phi$ is absorbed into the model. Estimating $T$ and the landing densities independently exposes it and lets us quantify it through $\kappa$.

Because they are conditional configuration-space distributions, the landing densities also serve as training targets for a generative model. Sampling $x_{t+\tau}\sim q_i$ and learning $p_\theta(x\mid X_i)$ gives a decoder that propagates the reweighted dynamics directly in configuration space, producing structures consistent with both the inferred kinetics and the reweighted thermodynamic ensemble.

\subsubsection{A decoder model: direct KL training of diffusion models with MaxEnt landing densities}
We now introduce a generative extension of our BICePs-reweighted DeepMSM framework. While the previous stage resampled training configurations, our objective here is to generate entirely novel configurations that are simultaneously consistent with the learned thermodynamics and kinetics. This requires a generative model for pairs of configurations sampled from the corrected transition density.

Our strategy builds on the generative Markov state model proposed by Wu et al.\cite{wu2018deep}, who introduced a conditional generative network to sample new configurations from metastable states. Here, we extend this framework.  To train the generative model, we use the landing densities $q(x_{t+\tau}\mid x_t\in X_i; \boldsymbol{\pi})$ learned during the DeepMSM construction, which already accounts for systematic error originating from the force field. These corrected densities are used to construct both the rate matrix $K$ and to train the generative component. Instead of the energy-distance loss originally used by Wu et al., we seek to train a conditional generative diffusion model $\tilde x_{t+\tau} \sim p_\theta(x_{t+\tau} \mid x_{t} \in X_i)$ with parameters $\theta$ whose output distribution matches $q_i(x)$. This can be formulated as direct distribution matching via the objective
\begin{equation}
  \mathcal{L}_{\mathrm{KL}}(\theta) = \sum_i \pi_i \, D_{\mathrm{KL}}\!\left(q_i(x)\,\|\,p_\theta(x\mid X_i)\right),
\end{equation}
where $\pi_i$ are the stationary state populations. Minimization of this objective is equivalent (up to an additive constant) to minimization of the negative conditional log-likelihood $\mathbb{E}_{x\sim q_i}[-\log p_\theta(x_{t+\tau} \mid x_{t} \in X_i)]$, which in diffusion models is implemented via a denoising-score matching objective under a fixed forward noising process.

In diffusion models, the conditional distribution $p_\theta(x\mid X_i)$ is implicitly specified by a latent-variable construction rather than an explicit likelihood.\cite{ho2020denoising} A forward diffusion process progressively maps samples $x_0 \sim q_i(x)$ to a simple reference distribution through a Gaussian noising schedule, while a neural network parameterizes the corresponding time-reversed denoising dynamics conditioned on the source state $i$.

Following training of the diffusion model, we obtain the transition matrix using the rewiring trick \cite{wu2018deep}, evaluating the configuration-space integral with samples drawn from the generator:

\begin{equation}
  K^{\mathrm{gen}}_{ij}=\int p_\theta(x_{t+\tau}\mid x_t\in X_i)\, \chi_j(x_{t+\tau})\,\mathrm{d}x_{t+\tau}.
  \label{eq:rewire_trick}
\end{equation}
Rows are stochastic by construction, since each row averages membership vectors that themselves sum to one.

In this work, we use this framework not as a generic data-driven density estimator, but as a direct variational approximation to the previously learned MaxEnt landing densities $q_i(x_{t+\tau})$. The diffusion model is not tasked with learning the time-ordered dynamics—that has already been achieved. Instead, it provides a conditional generative realization of the MaxEnt landing densities that define the transition model.

\section{Methods}
We first  validate our approach using a quadruple-well toy system, and then apply it to alanine dipeptide.  All neural network code is implemented using PyTorch\cite{paszke2017automatic}.

\subsection{Quadruple-well potential}
We validate the algorithm using a 1-D quadruple-well potential $U(x)=4\left(x^8+0.8 e^{-80 x^2}+0.2 e^{-80(x-0.5)^2}+0.5 e^{-40(x+0.5)^2}\right)$ as introduced by Prinz et al \cite{prinz2011markov}.
Training data were generated from a Brownian dynamics simulation in the quadruple-well potential using an Euler--Maruyama integration scheme
where $D=k_BT=1$, and configurations were saved every 100 integration steps, corresponding to an output interval of $\Delta t=0.01$ (same as Wu et al.\cite{wu2018deep}). The final dataset consisted of 500k saved configurations, and used a split 80\%/20\% for training and validation.

The $\chi$ network consists of four hidden layers of $64$ neurons each with ReLU activations. We warm-started $\chi$ by minimizing the VAMP2 score loss using a batch-size of 1000 unordered time-lag pairs of configurations with a lag time of $\tau=1$ during training. Then, both $\chi$ and $q$ were jointly trained using the negative log-likelihood loss with a learning rate of $1e^{-3}$.  Results used a time-lag of $\tau=5$ frames, which is when implied timescales begin to plateau.

To validate the performance of our approach to incorporate experimental information, we use the ensemble-average position, $<x>$, as a tunable experimental observable. As a reference case, to recapitulate the distribution of training data, the experimental observable was set equal to the average $x$ over all training data. This experimental observable was then modulated to investigate how dynamics change with respect to the experimental observable (see Results).

Both sets of eigenfunctions from the initial Koopman model and the BICePs-reweighted DeepMSM  are computed as $\psi_{k}(x) = \sum_{j} v_{k j} \chi_{j}(x)$, where $v_{k }$ are the right eigenvectors of the transition matrix $K$. To assess the agreement between BICePs-reweighted DeepMSM eigenfunctions $\psi_{k}$ and true eigenfunctions, we constructed a 100-state Markov transition operator on a one-dimensional grid for a potential derived from the predicted stationary distribution, $\mu(x)$, using acceptance probabilities from the Metropolis criterion.  These reference results are denoted \textit{True: Analytical ($\mu$)}, in Figures \ref{fig:quad_well_exp_+0.5} and \ref{fig:quad_well_exp_avg}.

For methods that estimate relaxation timescales, the data generation and model estimation procedures were repeated five times, and the reported values correspond to the mean and standard deviation across independent trials.

To evaluate estimator performance under biased, non-equilibrium training data, we apply four estimators, differing only in covariance construction and propagator estimation: (i) a VAMPnet, using the non-reversible Koopman matrix; (ii) a Symmetrized VAMPnet, which directly symmetrizes the instantaneous and time-lagged covariances; (iii) a Reversible VAMPnet, which constructs a Koopman-reweighted reversible estimate using importance weights ($w=\chi^\top\mathbf{C}_{00}^{-1}\hat{\pi}$); and (iv) a DeepMSM, which uses the same reweighted covariances to construct the prior followed by the similarity transformation and orthogonal rotation steps as given in Theory Section \ref{sec:deepMSM}.
Performance is assessed using state populations, the three slowest implied timescales, eigenfunctions, and the minimum transition-matrix entry, the latter of which tests whether the estimated matrix is stochastic rather than merely reproducing the correct spectral properties.

Similar to Mardt et al \cite{mardt2020deep}, many short trajectories $\{10^2,3\times10^2,10^3,3\times10^3,10^4\}$ are initialized from a biased mixture of Gaussians with standard deviation $\sigma=0.15$ centered at $x=\{-0.75,-0.25,0.25,0.75\}$ with respective weights ($\{0.15,0.70,0.09,0.06\}$) and propagated for $11$ frames. Populations, timescales, convergence and breakdown analyses all use $\tau=10$. Eigenfunctions are estimated separately from a ($10^5$)-frame equilibrium trajectory and from $2 \times 10^4$ short off-equilibrium trajectories. All experiments are repeated across five independent simulation trials. Error bars denote $\pm2 \sigma$ across trials, and heatmap errors report the median across trials.

\subsection{Alanine dipeptide}

All-atom explicit solvent molecular dynamics simulations of alanine dipeptide\cite{nuske2017markov} were performed using the AMBER99SB-ildn\cite{nerenberg2011optimizing} force field and TIP3P water model . Trajectories were 250 ns in length, with frames saved every 1 ps.

The Cartesian coordinates $(x,y,z)$ for all (10) heavy atoms were used as input features.  Our model was trained using a lag time of $\tau=1$ ps, as done in previous studies\cite{wu2018deep}. The $\chi$ network consisted of three hidden layers of 100 neurons with ReLU activations.   For training $\chi$, we warm-started $\chi$ by minimizing the VAMP2 score loss. Then, both $\chi$ and $q$ were jointly trained using the negative log-likelihood loss with a batch size of 4k samples of unordered time-lag pairs.

\subsection{Diffusion model details}
The conditional diffusion decoder was implemented as a fully connected neural network consisting of six hidden layers with 128 neurons per layer and ReLU activations. A cosine noise schedule with 300 diffusion time steps was used together with the velocity ($v$) parameterization objective.  During training, source states were sampled according to their stationary populations ($\pi_i$), while target configurations were sampled from the corresponding MaxEnt landing-densities $q_i(x_{t+\tau})$.  The decoder was trained to learn the conditional distribution $p_\theta(x_{t+\tau}\mid X_i)$ from samples drawn according to the joint distribution $\pi_i q_i(x_{t+\tau})$ rather than from the unweighted empirical training data.  For each training sample, the source state identity was provided as a one-hot conditioning vector. Prior to diffusion, coordinates were transformed using state-specific affine normalization parameters estimated from the corresponding landing configurations and held fixed throughout training. The diffusion latent dimension was chosen equal to the output dimensionality ($D=30$), such that diffusion and denoising were performed directly in the state-normalized coordinate space without an additional low-dimensional bottleneck.

Training was performed with a batch size of 4000 and an initial learning rate of ($10^{-3}$) using the Adam optimizer with weight decay ($10^{-5}$). The learning rate was reduced by a factor of 0.5 after five consecutive epochs without improvement in the training loss, and training was terminated after ten consecutive epochs without improvement. In all cases convergence was achieved in fewer than 100 epochs. On a MacBook M1 Pro, training required approximately 8 min, while generation of a trajectory containing ($2.5\times10^5$) configurations required approximately 1 min. Additional computational runtimes are reported in Table \ref{tab:runtime_evals}.


As an independent kinetic reference, we constructed a 100-state MSM from the same trajectory data. Conformations were represented by backbone torsions and discretized into 100 microstates by $k$-means clustering. Discrete trajectories were kept separate for each continuous trajectory so that no transitions were counted across trajectory boundaries. The active set was defined as the largest strongly connected set of the count matrix at the longest lag time considered and held fixed across all lag times, so that the implied timescale curves are directly comparable. At each lag time, transition counts from the active set were used to construct a reversible transition matrix using a maximum likelihood estimator\cite{prinz2011markov}. Reweighted populations were then imposed on this MSM by MaxCal, as in Eq. \ref{eq:maxcal}, providing implied timescales that can be compared directly with those of the reweighted DeepMSM.  Implied timescales for the generative model were obtained by the rewiring trick \cite{wu2018deep}: for each metastable state,  $10^{5}$ configurations were drawn from the trained decoder and re-encoded with the membership functions (Eq. \ref{eq:rewire_trick}). All timescales were averaged over five independent estimates.

\subsection{Validation of frozen membership functions}
Because the memberships $\chi$ are fixed after training, we took several steps to assess the quality of the learned state decomposition. We obtain near-crisp memberships with $\kappa=0.93$ for the quadruple-well system and $0.85$ for alanine dipeptide, corresponding to mean peak memberships of $0.95$ and $0.87$.

Reweighting changes the ensemble on which $\chi$ was optimized, so we next ask whether the encoder must be retrained. We recomputed the VAMP-2 score in the frozen basis under the reweighted frame weights $\mu$ and compared it with the score in an enriched basis containing $\chi$ and all pairwise products $\chi_i\chi_j$. Because the enriched basis contains $\mathrm{span}\{\chi\}$, it provides an upper bound on the improvement attainable by re-optimizing the encoder on the same features. The difference in scores increased only from $0.001$ to $0.006$ for the quadruple well and from $0.029$ to at most $0.038$ for alanine dipeptide across the J-coupling and dihedral-angle restraints. Thus, the frozen memberships remain within one percent of the attainable optimum after reweighting, eliminating the need for retraining when new experimental information is imposed.  This result confirms that frozen memberships are able to accurately resolve the slow dynamics, and that they are essentially preserved after reweighting.

\section{Results}

Here, we construct a general framework for learning both thermodynamics and kinetics that agrees with experimental measurements while considering all sources of uncertainty. We demonstrate how systematic error in molecular dynamic simulations, which we use as training data, can be corrected by applying experimental restraints to reshape both the landing densities $q(x_{t+\tau})$ and stationary distribution $\mu(x_{t+\tau})$ under the principle of maximum entropy. Our results are validated by comparing the thermodynamics and kinetics obtained from our framework against those obtained from (i) the prior model, (ii)  Maximum Caliber-reweighted MSMs, and (iii) a Metropolis-style Markov transition operator for the 1-D toy system.  Throughout the remainder of this work, we use the term  \textit{prior} to refer to the Koopman/Markov model constructed using simulated training data alone, without experimental restraints.

In this section,  we first build DeepMSMs of Brownian dynamics simulations in a 1-D quadruple-well potential, subject to different experimental restraints of increasing severity. Next, we apply our framework to all-atom molecular dynamics simulations of alanine dipeptide\cite{nuske2017markov}, and demonstrate the use of different types of time-averaged experimental observables.  Finally, we use the alanine dipeptide system to demonstrate the use of generative molecular modeling by training a diffusion model to mimic the MaxEnt landing densities conditioned on BICePs-reweighted populations. This produces trajectories of novel configurations that agree with the experimental data. Furthermore, we demonstrate the ability of our model to generate physically valid configurations even in the absence of many training samples (left-handed $\alpha$-helix region: $\phi=+60^\circ$).

\subsection{Quadruple-well potential}
We use a quadruple-well toy system in which the "experimental" observable is the ensemble-average position $\langle x \rangle$.  The model was constructed using four states and a lag time of $\tau=5$, following previous studies \cite{wu2018deep}. The experimental observable was initially chosen to be the ensemble-averaged position over the training data ($\langle x \rangle^{\text{Exp.}} = 0.1$).

First, as a consistency check, we  considered the case in which the experimental observable exactly matches the training-data average, $\langle x\rangle^{\mathrm{Exp.}}=0.10$. In this case, the reweighted stationary distribution is indistinguishable from the prior distribution (Figure~\ref{fig:quad_well_exp_avg}a), indicating that the MaxEnt reweighting agrees strongly with the prior populations and introduces no spurious perturbations when the experimental information is already satisfied by the simulation ensemble. The corresponding state memberships and landing densities are shown in Figure \ref{fig:quad_well_exp_avg}b-c.  The relaxation timescales and eigenfunctions ($\psi_{k}$) all remain in close agreement with the original model (Figure \ref{fig:quad_well_exp_avg}d-e). These results demonstrate that the reweighting procedure preserves both thermodynamic and kinetic properties when no correction to the simulation ensemble is required.

\begin{figure}[h!]
\centering
\includegraphics[width=\linewidth]{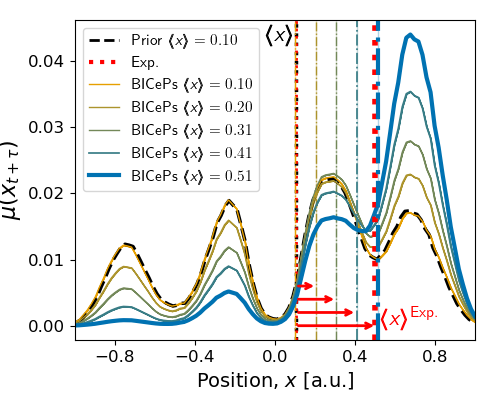}
  \caption{
    \small Response of the stationary distribution $\mu$ to changes in the experimental observable. The prior stationary distribution (black dashed curve) has an average position of $\langle x \rangle = 0.1$.  Increasing the experimental value from $\langle x \rangle^{Exp.} = 0.1$ to $0.5$ progressively shifts probability density toward larger values of $x$.   The resulting stationary distributions satisfy the imposed experimental averages while remaining consistent with the kinetic constraints learned from simulation data. Vertical dashed lines indicate the target experimental values, and solid curves show the corresponding reweighted stationary distributions.
  }
\label{fig:quad_well_mu_exp_scan}
\end{figure}

Next, to examine how the model responds to experimental restraints that increasingly deviate from the prior, we varied the target observable over the range $\langle x\rangle^{\mathrm{Exp.}}=\{0.1,0.2,0.3,0.4,0.5\}$  (Figure \ref{fig:quad_well_mu_exp_scan} ). As the target average position increases, the stationary probability density is progressively redistributed toward larger values of x, with the largest changes occurring in the right-most metastable basin. Importantly, the resulting distributions evolve smoothly as a function of the imposed experimental observable.

\begin{figure*}[!htbp]
\centering
\includegraphics[width=\linewidth]{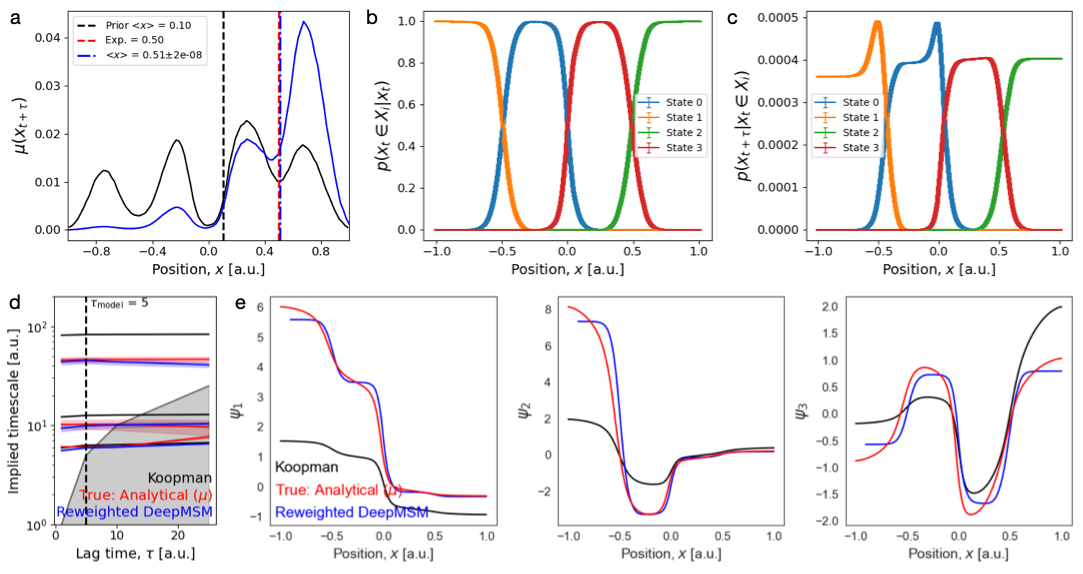}
  \caption{\small Performance of BICePs-reweighted DeepMSMs for estimating the kinetics and thermodynamics when the experimental data ($<x>=0.50$) differs from the training data ($<x>=0.10$).  (a) The stationary distribution with a predicted average observable of $<x>=0.51\pm2x10^{-8}$. (b) The state-membership probabilities. (c) The landing densities. (d) Relaxation timescales plateau across lag times $\tau$. The BICePs-reweighted DeepMSM timescales have strong agreement with the True timescales, which are much faster than that of the prior. (e) The corresponding non-trivial eigenfunctions $\psi_{k}$ for each of the slow modes.
  }
\label{fig:quad_well_exp_+0.5}
\end{figure*}

We next discuss the results for the case $\langle x\rangle^{\mathrm{Exp.}}=0.50$.   The resulting stationary distribution shifts populations toward the right-most basin, yielding a significantly different equilibrium landscape (Figure \ref{fig:quad_well_exp_+0.5}a). The MaxEnt landing densities shown in Figure \ref{fig:quad_well_exp_+0.5}c reflect the modified equilibrium ensemble, illustrating an increase in probability density towards configurations with larger  $x$ values (compare with Figure \ref{fig:quad_well_exp_avg}c).

The altered stationary distribution produces significantly faster relaxation dynamics than those of the Koopman model built from only training data. Figure \ref{fig:quad_well_exp_+0.5}d shows that the implied timescales predicted by the BICePs-reweighted DeepMSM closely match those obtained from the analytical reference model (truth) constructed from the reweighted stationary distribution, while differing substantially from the prior dynamics. This agreement indicates that the method correctly propagates thermodynamic reweighting into the kinetic model rather than merely reproducing the target equilibrium distribution.

Consistent with this interpretation, the leading non-trivial eigenfunctions closely overlap with those obtained from the analytical reference model (Figure \ref{fig:quad_well_exp_+0.5}e), demonstrating that both the slow dynamical modes and their associated timescales are accurately recovered after reweighting.

Up until this point, all of the presented results have used a single long equilibrium trajectory. We now turn to the case that motivates the Koopman reweighting (Eqs. \ref{eq:covariances} and \ref{eq:koopman}): many short trajectories started from an off-equilibrium distribution.  For this case, we compare the DeepMSM against a non-reversible VAMPnet, a symmetrized VAMPnet, and a reversible VAMPnet. Given abundant equilibrium data all four estimators recover the reference populations, timescales, and slow eigenfunctions (Figures \ref{fig:method_compare_pops_timescale}a,c and \ref{fig:method_compare_eigenfunctions}a--c). Given off-equilibrium data, the symmetrized estimator is biased in every observable, underestimating the slowest timescale and distorting $\psi_2$, while the other three remain unbiased (Figures \ref{fig:method_compare_pops_timescale}b,d and \ref{fig:method_compare_eigenfunctions}d--f). The DeepMSM and the reversible VAMPnet share the same slow eigenfunctions, as they are built from the same reweighted covariances.

Figure \ref{fig:method_compare} shows that the symmetrized estimator sits on a bias floor that additional sampling does not remove, while the remaining three estimators converge as the number of trajectories grows. Only the DeepMSM's transition matrix is nonnegative at every sample size, so it alone is unbiased given off-equilibrium data and therefore preferred wherever individual transition probabilities are needed. The DeepMSM has two failure modes (Figure \ref{fig:deepmsm_crispness}). Too few trajectories degrades both populations and timescales, and is curable by more sampling. Insufficient membership crispness instead leaves a floor in the populations that sampling cannot remove, since overlapping memberships cannot resolve metastable populations however well the propagator is estimated. The timescales are far less sensitive to crispness, consistent with the eigenvalues of $K$ depending on the membership subspace rather than on the individual membership functions.

\subsection{Alanine dipeptide}
To evaluate the performance of our framework on a realistic molecular system, we reweighted DeepMSMs built from molecular dynamics (MD) simulations of alanine dipeptide. Because of its small size, well-characterized conformational landscape, and extensive use as a benchmark system for molecular kinetics, the alanine dipeptide system provides an ideal test case. We consider two classes of observables: backbone dihedral angles and J-couplings derived from the $\phi$ torsion. Using our MaxEnt-reweighted landing densities, we train a diffusion-based model for each test case.

\subsubsection*{Dihedral angles $\phi$ and $\psi$ as experimental observables.}

As a preliminary demonstration, we first consider a simplified scenario in which the backbone dihedral angles $\phi$ and $\psi$ are treated as experimental observables (Figure \ref{fig:reweighted_dynamics_angles}). Because these observables directly define the conformational landscape, they provide an intuitive test of whether the model can recover a prescribed equilibrium ensemble and the corresponding kinetics.

Applying experimental data restraints corresponding to $(\phi,\psi)=(-74^\circ,-12^\circ)$ redistributes the population density within the Ramachandran landscape, increasing the population of state 4—the targeted basin (Figure \ref{fig:reweighted_dynamics_angles}a). The reweighted ensemble accurately reproduces the experimental observables (Figure \ref{fig:reweighted_dynamics_angles}c) and yields relaxation timescales that differ from those of the Koopman model while remaining in good agreement with 100-state MSM + MaxCal estimates across lag times (Figure \ref{fig:reweighted_dynamics_angles}b).  Furthermore, the DeepGenMSM is also kinetically consistent with the reweighted DeepMSM. These results demonstrate that changes in equilibrium populations are propagated into a self-consistent kinetic model rather than being treated as a purely thermodynamic reweighting, and thus can be used to condition diffusion models for generative dynamics.

To evaluate whether the diffusion model trained on the landing densities (Figure \ref{fig:landing_densities_angles}) can generate physically realistic configurations consistent with the reweighted ensemble, we generated trajectories from the BICePs-reweighted DeepGenMSM. The distributions of bond lengths and bond angles remain largely unchanged relative to the training data, indicating that local molecular geometry is preserved (Figures \ref{fig:gen_phi_psi_angles}a,b). The largest deviations occur for internal coordinates directly adjacent to the restrained $\phi$ and $\psi$ dihedrals, consistent with the intended redistribution of conformational populations. Furthermore, the generated trajectories accurately reproduce the reweighted stationary distributions along both $\phi$ and $\psi$, as well as the full two-dimensional $(\phi,\psi)$ landscape (Figures \ref{fig:gen_phi_psi_angles}c–e), demonstrating that the diffusion model successfully mimics the thermodynamics and kinetics encoded by the MaxEnt landing densities.

\subsubsection*{$J$-coupling constants as experimental observables.}

\begin{figure*}[!htbp]
\centering
\includegraphics[width=\linewidth]{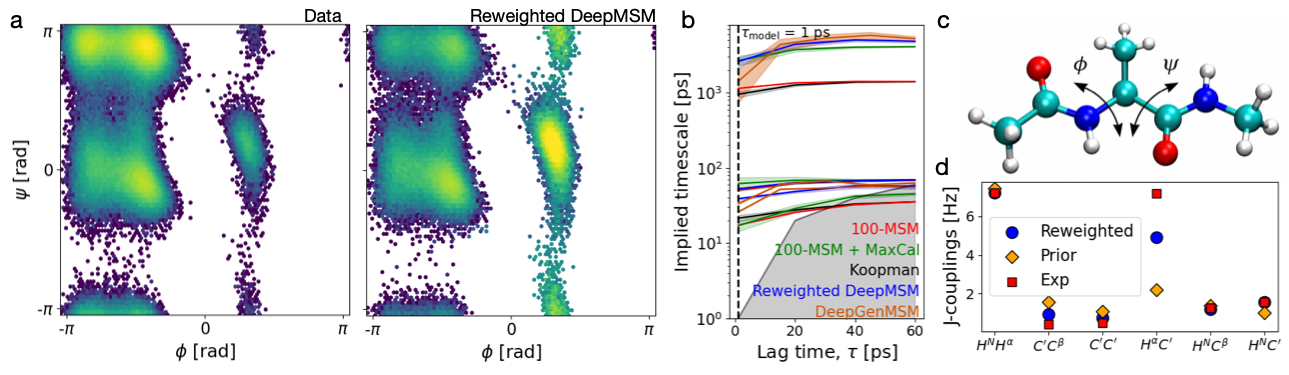}
  \caption{\small Performance of the BICePs-reweighted DeepMSM in estimating the thermodynamics and kinetics of alanine dipeptide when given six experimental J-couplings ($\phi=60^\circ$). (a) Training data distribution and reweighted stationary distribution. (b) Relaxation timescales from the models built using only training data (Koopman and the 100-state MSM), the 100-state MSM + MaxCal reweighting, the reweighted DeepMSM, and the DeepGenMSM. (c) Comparison of ensemble-averaged observables from the prior, BICePs-reweighted DeepMSM, and experimental data.
  }
\label{fig:reweighted_dynamics_J_60_deg}
\end{figure*}

First, we consider the case in which the experimental $J$-couplings are set equal to their ensemble-averaged values computed from the training data (Figure \ref{fig:reweighted_dynamics_J_exp_avg}). In this regime, the experimental information is fully consistent with the simulation ensemble, providing a stringent test that the reweighting procedure does not introduce spurious thermodynamic or kinetic perturbations. As expected, the reweighted stationary distribution closely overlaps the prior distribution, predicted $J$-couplings agree with both the experimental targets and the values computed from the original simulation data, and the relaxation timescales closely match those of the Koopman model. Consistent with these results, trajectories generated from the diffusion model that was trained on our learned landing densities (Figure \ref{fig:landing_densities_avg}) preserve the original bond-length and bond-angle distributions and accurately reproduce the stationary distributions along both $\phi$ and $\psi$, as well as the full two-dimensional Ramachandran landscape (Figure \ref{fig:gen_J_exp_avg}). Together, these results demonstrate that when the experimental observables are already satisfied by the simulation data, the framework faithfully recovers both the thermodynamics and kinetics of the original model.

A comparison of our MaxEnt reweighted landing densities to those of a DeepMSM\cite{wu2018deep} trained on only trajectory data (no experimental restraints) using the negative log-likelihood loss show remarkably similar results (Figures \ref{fig:landing_densities_avg} \& \ref{fig:landing_densities_avg_deepMSM_prior}).

Next, we consider a more demanding scenario in which the experimental observables correspond to six $J$-coupling constants computed from a specific $\phi=60^\circ$ conformation. This region lies within the left-handed $\alpha$-helical basin and is only sparsely sampled in the training data (only $\sim$1k frames belong to state 1). Consequently, satisfying these experimental restraints requires a substantial redistribution of population density away from the original simulation ensemble.

The resulting reweighted stationary distribution is shown in Figure \ref{fig:reweighted_dynamics_J_60_deg}a. Relative to the prior model, population density is transferred from the dominant $\beta$-sheet and right-handed $\alpha$-helical regions into the left-handed $\alpha$-helical basin near $\phi \approx 60^\circ$ — consistent with the experimental restraints. The reweighted ensemble produces J-coupling values that move substantially toward the experimental targets (Figure \ref{fig:reweighted_dynamics_J_60_deg}c).

Importantly, the thermodynamic perturbation is accompanied by a corresponding change in the kinetics. The relaxation timescales predicted by both the reweighted DeepMSM and DeepGenMSM differ significantly from those of the Koopman model and closely track the 100-state MSM + MaxCal reference estimates across lag times (Figure \ref{fig:reweighted_dynamics_J_60_deg}b).

At early lag times, the deep generative model's kinetic accuracy is limited by the effective sample size of the reweighted landing densities, but still within error of the rweighted DeepMSM and Koopman estimates. For states whose reweighted density concentrates on a small fraction of the initial simulated ensemble (here state 1, $N_{\text{eff}} = 1.2  \times 10^3$ frames), the decoder has insufficient training data and the corresponding transition probabilities (Eq. \ref{eq:rewire_trick}) are systematically underestimated. At longer lag times, the dispersion in landing density grows, increasing $N_{\text{eff}}$ for states 1 and 3.

As expected, we found strong overlap between the generative model landing densities (Figure \ref{fig:gen_landing_densities_J_60_deg}) and the MaxEnt landing densities $q$ used as training data for this model, with the largest JSD being 0.083 (state 1). For this analysis, a histogram was constructed from 100k generated samples from each conditional state and directly compared to $q$.

\begin{figure*}[!htbp]
\centering
\includegraphics[width=\linewidth]{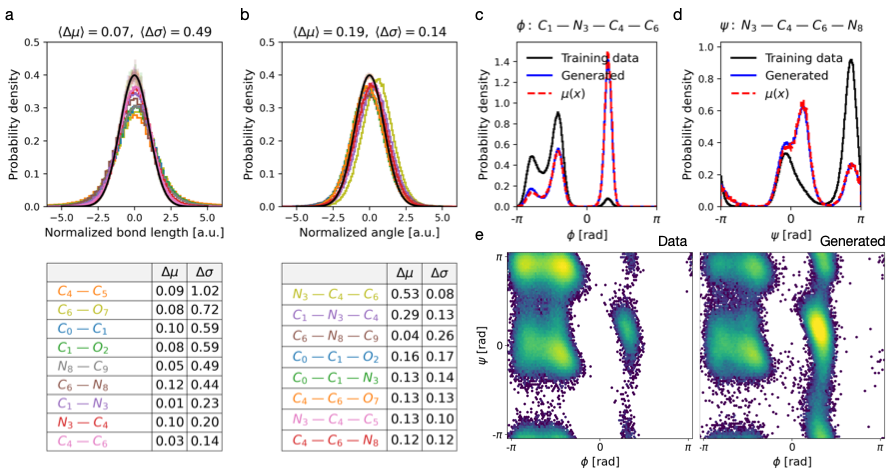}
  \caption{\small Performance of BICePs-reweighted Deep Generative MSM in estimating the thermodynamics and kinetics of alanine dipeptide when given six experimental J-couplings ($\phi=60^\circ$).  (a-b) Normalized bond-length and angle distributions from training data (black curve) and from the trajectory generated from the model. Tables report the differences in the mean, $\Delta \mu$ and standard deviation, $\Delta \sigma$ between the training data and generated trajectory. (c-e) Stationary distributions of $\phi$ and $\psi$. The stationary distribution $\mu$ (red curve) is overlaid to show agreement with the density from the generated trajectory.
  }
\label{fig:gen_J_60}
\end{figure*}

To evaluate whether the learned diffusion model can generate physically realistic configurations consistent with the reweighted dynamics and landing densities (Figure \ref{fig:landing_densities_J_60_deg}), we generated novel trajectories from the BICePs-reweighted DeepGenMSM. Bond-length and bond-angle distributions remain largely unchanged relative to the training data (Figures \ref{fig:gen_J_60}a,b), indicating that local molecular geometry is preserved. Notably, the largest deviations occur for bond angles and bond lengths directly adjacent to the restrained $\phi$ torsion, including $N_3\!-\!C_4\!-\!C_6$ and $C_1\!-\!N_3\!-\!C_4$. These localized structural changes are consistent with the experimentally induced redistribution of population toward the $\phi=60^\circ$ region of conformational space.

The generated trajectories accurately reproduce the stationary distributions predicted by the reweighted DeepMSM along both $\phi$ and $\psi$ (Figure \ref{fig:gen_J_60}c,d), with the generated densities nearly indistinguishable from the target stationary distribution $\mu(x)$. Furthermore, the two-dimensional ($\phi,\psi$) distribution obtained from the generated trajectory closely matches the reweighted equilibrium ensemble (Figure \ref{fig:gen_J_60}e). Together with the low Jensen-Shannon divergences between the learned generative model $p_\theta$ and the target landing densities $q$, these results demonstrate that the diffusion model successfully learns the state-conditioned landing densities while generating continuous, physically realistic trajectories that explore previously under-sampled regions of conformational space.

Together, these results demonstrate that experimentally derived observables can be used not only to reweight equilibrium populations, but also to infer the corresponding kinetic model and generate physically realistic molecular configurations consistent with the resulting thermodynamic ensemble.

\section{Discussion}

\subsection{DeepMSM comparison with restraints applied \emph{during} training.}
\label{sec:mardt-comparison}

To distinguish our approach from that of Mardt et al., we refer to the method of Mardt \emph{et al.} as the \emph{training-constrained DeepMSM}, reflecting that physical constraints and experimental information are incorporated during end-to-end optimization. Their approach augments the training objective with restraint terms,
\begin{equation}
\mathcal{L}_{\mathrm{tot}} =  \mathcal{L}_{\mathrm{MSM}} + \sum_i \lambda_i \left\lVert O_i-\mathbb{E}[a_i]\right\rVert^2,
\label{eq:mardt-loss}
\end{equation}
while enforcing nonnegativity and reversibility by constraining the trainable parameters \cite{mardt2020deep,mardt2021progress}. At the operator level, the two approaches are closely related: the symmetric coupling matrix of Mardt et al. and our whitened propagator $\mathbf S_0$ represent the same reversible operator under different scalings. The principal distinction is therefore not the learned operator itself, but where physical constraints and experimental information are imposed.

In the training-constrained DeepMSM, experimental observables are balanced against the kinetic objective through the restraint weights $\lambda_i$, whose values must be specified \emph{a priori}. These hyperparameters determine the tradeoff between reproducing the experimental observables and preserving the learned kinetics, yet no principled choice is generally available. In our DeepMSM, the target stationary distribution enters as a hard constraint through the Maximum Caliber optimization, while the uncertainty in the experimental data is treated upstream within the Bayesian framework of BICePs. Rather than selecting restraint strengths, BICePs infers the corresponding uncertainty parameters directly from the data by marginalizing over them, producing a posterior distribution over $\bm\pi$. The DeepMSM therefore contains no restraint hyperparameters, and uncertainty is propagated naturally from the experimental measurements into the kinetic model.

A second distinction is that our DeepMSM separates representation learning from experimental reweighting. Once the memberships have been learned, the prior transition matrix is fixed, and any target stationary distribution can be incorporated through a single convex MaxCal solve without retraining the neural network. By contrast, because experimental restraints are embedded directly in the training objective, changes to the restraints, their weights, or the training conditions generally require a new end-to-end optimization in the training-constrained DeepMSM.

The two approaches also differ in how equilibrium information influences kinetics. Mardt \emph{et al.} observed that restraining equilibrium observables primarily improved thermodynamic quantities, while additional kinetic restraints were required to substantially improve the relaxation timescales. Structurally, matching equilibrium expectation values through a loss function does not, by itself, require the learned propagator to change. In our DeepMSM, the stationary distribution is imposed directly on the equilibrium flux through Eq. \ref{eq:maxcal}, so the transition matrix is modified by construction through Eq. \ref{eq:Treweighted}. Because MaxCal operates directly on the equilibrium flux, changes in the stationary populations necessarily induce corresponding changes in the transition rates, consistent with the linear free-energy interpretation discussed above.

Finally, our DeepMSM separates the construction of a valid Markov model from the incorporation of experimental information. The orthogonal transformation used to construct the prior transition matrix preserves the VAC eigenspectrum exactly while enforcing reversibility, normalization, and nonnegativity. Experimental populations are introduced only afterward through the MaxCal reweighting, ensuring that changes in the kinetics arise solely from the experimentally informed stationary distribution rather than from enforcing physical constraints during training.

The two approaches are therefore complementary. The training-constrained DeepMSM incorporates physical constraints and experimental information directly during representation learning, whereas our DeepMSM first constructs a spectrally faithful, physically admissible prior and subsequently incorporates experimental populations through a convex Maximum Caliber reweighting. This separation preserves the learned VAC spectrum prior to reweighting, avoids retraining when new equilibrium information becomes available, and naturally supports posterior uncertainty propagation through the MaxCal solution (Section \ref{sec:appendix_theory}).

\subsection{Caveats of this method.}

A fundamental limitation of the present framework is that the inferred kinetics are constrained by MaxCal through equilibrium populations. By combining BICePs reweighting with MaxEnt/MaxCal principles, we obtain the minimally perturbed kinetic model consistent with the experimentally refined equilibrium ensemble. While detailed balance couples thermodynamics and kinetics, time-averaged observables do not uniquely determine the underlying dynamics. Consequently, the resulting kinetics should be interpreted as the least-biased dynamics consistent with both the simulation and available experimental constraints, rather than an exact reconstruction of the true molecular kinetics. Obtaining a more accurate kinetic model would require implementing time-dependent experimental observables \cite{henning2025bayesian, kolloff2023rescuing}, which remain relatively scarce for most biomolecular systems.


A second caveat concerns the construction of the underlying DeepMSM. While the Bayesian Inference of Conformational Populations (BICePs) algorithm has no direct dependency on system size or complexity, the reweighted DeepMSM (and our DeepMSM prior) critically depends on the learned state membership probabilities. Our approach leverages pre-trained Koopman/Markov models together with experimental information to determine how the underlying dynamics should be adjusted to achieve consistency with experimental observables. While the method performs well for the toy systems and alanine dipeptide studied here, both VAMPnets and DeepMSMs may encounter challenges in more complex biomolecular systems where non-Markovian effects cannot be neglected, such as systems exhibiting allostery, parallel pathways, or hidden slow modes. Furthermore, the optimal number of states may become increasingly difficult to determine as system complexity grows, potentially impacting both the accuracy and interpretability of the resulting kinetic model. Although these issues were not significant for the systems considered in this work, future applications to larger and more complex biomolecular systems may require more sophisticated approaches for state construction and model training.

\subsubsection*{From kinetics to state decomposition: determining the optimal number of states}
A central challenge in constructing Markov state models is determining the optimal number of states that faithfully represent the underlying kinetics without imposing unnecessary state decomposition. Methods such as VAMPnets implicitly assume a fixed-dimensional Koopman-invariant subspace, requiring the number of states $m$ to be specified a priori, which can bias the learned representation toward a predefined kinetic resolution rather than revealing the intrinsic dynamical structure. In complex systems, where timescale separation may be weak or hidden, this choice becomes critical, as an inappropriate $m$ can lead to either over-coarse models that obscure slow processes or overly fine models that fragment metastable states\cite{liu2026unveiling}. A more principled perspective prioritizes the identification of the slow kinetic modes first, and treats the optimal number of states as a consequence of this analysis rather than an input parameter. Spectral gap approaches provide a natural criterion for identifying the number of dominant slow processes by examining separations in the eigen/singular value spectrum \cite{tiwary2016spectral}.

Building on this idea, we propose a general framework for analyzing complex systems in which the kinetic structure is determined prior to state decomposition. In practice, this begins by constructing a high-resolution prior model—either via a VAMPnet or MEMnets\cite{liu2025memory} architecture—using a sufficiently large number of microstates to capture the relevant dynamical variability. From this model, the slow kinetic modes are identified through spectral analysis, and the system is subsequently coarse-grained into an optimal number of macrostates based on the observed spectral gaps. Our reweighing framework naturally follows by providing a fully admissible DeepMSM prior that can be used for reweighting, while preserving the spectrum of the coarse-grained model.  In this framework, the number of states is not imposed but emerges naturally from the hierarchy of dynamical timescales, making it particularly well-suited for complex systems where both memory effects and limited sampling are present.

\subsection{Generative models with explicit data-restraint conditioning.}
In our framework, experimental information is incorporated indirectly through the BICePs-reweighted stationary distribution, and the diffusion model is subsequently trained to reproduce the MaxEnt landing densities. Despite the absence of explicit restraint conditioning, the resulting model generates conformations consistent with the experimentally informed populations while maintaining the local structural characteristics of the training data. As shown in Figure \ref{fig:gen_J_60}, bond-length, bond-angle, and torsional distributions remain in close agreement with the reference ensemble, while the model is still able to generate novel configurations in sparsely sampled regions of conformational space, such as the left-handed $\alpha$-helical basin near $\phi \approx 60^\circ$. This suggests that population-level reweighting is sufficient to transfer the experimental information to the generative model without introducing distortions to local molecular geometry.

An alternative strategy would be to condition the diffusion model directly on the experimental observables, as recently proposed through minimum-excess-work guidance\cite{kolloff2026minimum} and related approaches.
Such methods offer the advantage of incorporating experimental information directly during sampling, without first constructing a reweighted ensemble.  However, direct conditioning also introduces the risk of overemphasizing sparse experimental restraints at the expense of structural realism, potentially requiring additional regularization or physical priors to preserve bond, angle, and torsional distributions. Furthermore, the computational cost of observable-guided sampling likely includes additional overhead due to optimization, determination of guidance parameters, and possibly heavier loss functions, whereas our approach only requires MaxEnt reweighting followed by standard diffusion-model training to mimic the rewighted landing densites.  Consequently, for the present problem, conditioning through the reweighted stationary distribution provides a simple, computationally efficient, and physically robust means of incorporating experimental information, while the potential advantages of direct observable conditioning remain unclear, but an interesting direction for future investigation.

\section{Conclusion}
In this work, we introduced BICePs-reweighted Reversible DeepMSMs, a unified framework for integrating experimental observables into both thermodynamic and kinetic models of molecular systems. Given a dynamical model that encodes trajectory information into state memberships---VAC, VAMPnets, DeepMSM, and related approaches---an orthogonal transformation applied post hoc in the whitened basis converts the Koopman estimate into a reversible DeepMSM prior, preserving the learned VAC spectrum exactly and requiring no retraining of the network. Combining this prior with BICePs and MaxEnt/MaxCal principles, the method refines stationary populations, transition dynamics, and state-conditioned landing densities while rigorously accounting for uncertainty, and permits reweighting to new equilibrium information without retraining. Using a quadruple-well toy system, we demonstrated that perturbations to ensemble-averaged observables produce predictable changes in both thermodynamics and kinetics that closely match analytical and MaxCal reference models. We then applied the framework to alanine dipeptide, where experimentally restrained backbone dihedral angles and J-coupling constants similarly induced physically interpretable changes in the equilibrium ensemble and associated relaxation processes. Finally, we showed that the learned MaxEnt landing densities can be used to train diffusion-based generative models capable of producing physically realistic molecular trajectories that satisfy the imposed experimental restraints while exploring previously under-sampled regions of configuration space. Overall, these results establish a general strategy for incorporating time-averaged experimental information into kinetic molecular models and suggest a path toward experimentally grounded generative simulations capable of simultaneously reproducing molecular structure, thermodynamics, and dynamics.


\section*{Conflicts of interest}
Authors declare no conflicts of interest.

\section*{Data and Software}
The BICePs algorithm is openly available at \href{https://github.com/vvoelz/biceps}{github.com/vvoelz/biceps}. All calculations in this work were performed using the development version biceps\_v3.0a. The specific version can be accessed at: \href{https://github.com/vvoelz/biceps/tree/biceps_v3.0a}{github.com/vvoelz/biceps/tree/biceps\_v3.0a}.
All scripts, Jupyter notebook examples and analysis can be found here. Detailed documentation and tutorials can be found here: \href{https://biceps.readthedocs.io}{biceps.readthedocs.io}. For any issues or questions, please submit the request on GitHub.

\section*{Acknowledgements}
RMR, and VAV are supported by National Institutes of Health grants R01GM123296 and R35GM163789.

\bibliography{references}

\appendix

\label{sec:appendix_theory}

\subsubsection{Uncertainty propagates without additional parameters.}
The construction introduces no additional fitting parameters beyond the target stationary distribution. Once the prior flux $F^{(0)}$ is fixed, the reweighted transition matrix is a deterministic map $T=\Phi(\bm\pi;\mathbf C_{00},\mathbf C_{01})$, so uncertainty in the BICePs posterior over $\bm\pi$ propagates directly to uncertainty in the kinetics. Posterior samples of $\bm\pi$ therefore induce posterior distributions over implied timescales, committors, reactive fluxes, and mean first-passage times through repeated Sinkhorn solves. Because the Sinkhorn fixed point $\alpha\circ(F^{(0)}\alpha)=\bm\pi$ is differentiable,
\begin{equation}
\frac{d\bm\alpha}{d\bm\pi} = \left[\mathbf D_{F^{(0)}\alpha}+\mathbf D_\alpha F^{(0)}\right]^{-1},
  \label{eq:uncertainties}
\end{equation}
from which gradients of kinetic observables follow directly.

%
%
%


\newpage
\clearpage

\onecolumngrid
\setcounter{page}{1}
\renewcommand{\thepage}{S\arabic{page}}

\setcounter{figure}{0}
\renewcommand{\thefigure}{S\arabic{figure}}

\setcounter{table}{0}
\renewcommand{\thetable}{S\arabic{table}}

\setcounter{equation}{0}
\renewcommand{\theequation}{S\arabic{equation}}

\begin{center}
{\Large\bfseries Supplemental Material}

\vspace{1cm}

{\large\bf \mytitle}

\vspace{0.5cm}

Robert M. Raddi and Vincent A. Voelz

\vspace{0.25cm}

Department of Chemistry, Temple University,
Philadelphia, Pennsylvania 19122, USA

\end{center}

\vspace{1cm}

This PDF contains additional figures, tables, and computational details
supporting the results presented in the main text.

\bigskip
\newpage

\subsection*{Details of parameters used in BICePs posterior sampling}

\subsubsection*{General BICePs reweighting}
\label{sec:reweighting_details}
The range of sampled uncertainty parameters $\sigma$ on a grid of logarithmically-spaced values between 0.001 to 200, to enforce the Jeffrey's prior.   Each grid value in the list was a factor of 1.02 larger than the next: [\texttt{1.00e-03 1.02e-03 1.04e-03 1.06e-03, ..., 1.95e+02 1.98e+02}], resulting in a list of 617 values.  BICePs used 100 replicas for the quadruple-well system and 500 replicas for alanine dipeptide.  Populations were averaged over 10 independent Markov chains for each experiment. All MCMC simulations reached convergence within 100k steps.

\begin{table*}[t]
\centering
\caption{Estimated computational runtimes for NN training, model construction, and reweighting of DeepMSMs for both systems in this study. Measurements were performed on a MacBook M1 Pro, with values reported as the mean over five independent runs. Over 95\% of the elapsed time comes from NN training.}
\label{tab:runtime_evals}
\begin{tabular}{|c|c|l|c|}
\hline
  System & \# of states & (\# of) Experimental observables & Elapsed time [s]  \\
\hline
  Quadruple-well & 4 & (1) Average position, set equal to average over the training data: $<x>^{Exp} = <x>^{Prior}$ & 173 $\pm$ 6 \\
  Quadruple-well & 4 & (1) Average position,  $<x>^{Exp} = +0.50$ & 157 $\pm$ 13 \\
  Alanine dipeptide & 6 & (6) J-couplings set equal to the average over the training data & 272 $\pm$ 20  \\
  Alanine dipeptide & 6 & (6) J-couplings computed at $\phi=+60^\circ$ & 257 $\pm$ 21  \\
  Alanine dipeptide & 6 & (2) Dihedral angles $(\phi=-74^\circ$, $\psi=-12^\circ)$   & 308 $\pm$ 35  \\
\hline
\end{tabular}
\end{table*}

%
%
%
%

%

\begin{figure*}[htbp]
\centering
\includegraphics[width=\linewidth]{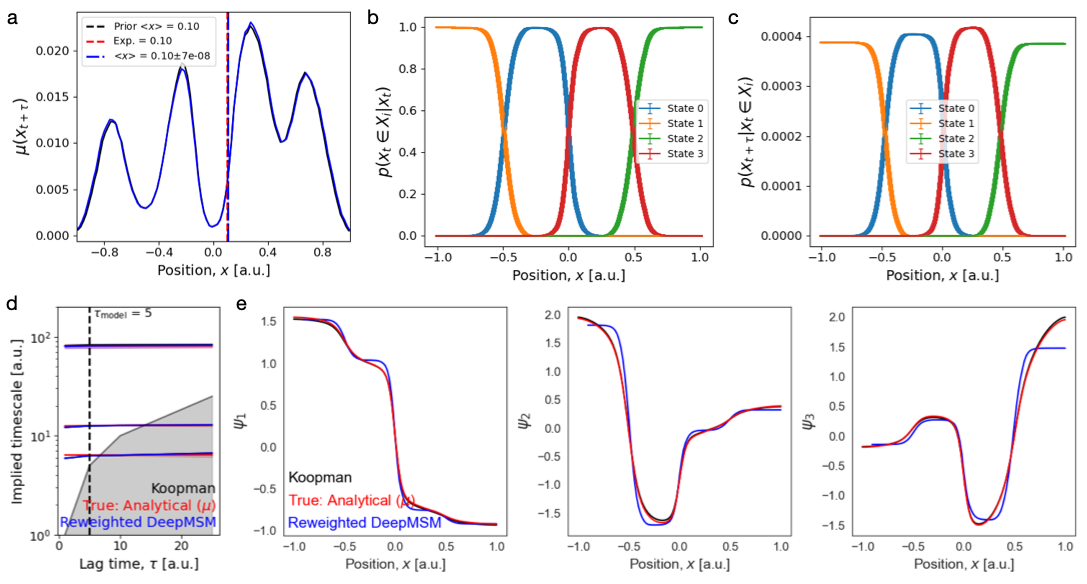}
  \caption{\small Performance of BICePs-reweighted DeepMSMs for recapitulating the kinetics and thermodynamics when the experimental data ($<x>=0.10$) agrees perfectly with the training data.  (a) The predicted stationary distribution (blue) overlaps very well with the prior (black) yielding a predicted average observable of $<x>=0.10\pm7.0x10^{-8}$. (b) The state-membership probabilities. (c) The landing densities. (d) Relaxation timescales at a lag time of $\tau=5$. BICePs-reweighted DeepMSM timescales very closely match the top slowest timescales of the Koopman model built from only using training data.
  (e) The non-trivial eigenfunctions strongly overlap.
  }
\label{fig:quad_well_exp_avg}
\end{figure*}

\begin{figure*}[htbp]
\centering
\includegraphics[width=\linewidth]{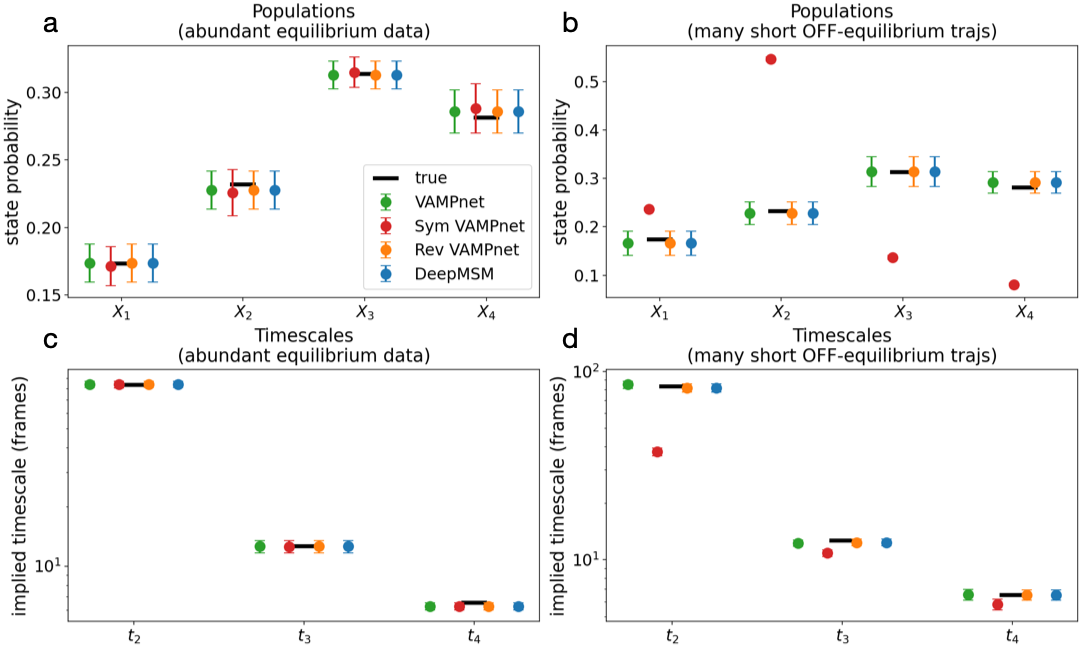}
  \caption{\small Populations and implied timescales for four estimators on the quadruple-well potential, using fixed memberships ($\kappa=0.88$) and lag $\tau=10$ frames. (a,b) Stationary probabilities of the four soft states. (c,d) The three slowest implied timescales ($t_k=-\tau/\ln\lambda_k$) in frames. The left column uses abundant equilibrium data and the right column uses many short trajectories started from a biased distribution. Black bars are the ground truth. Markers are means over five independent trials, with $\pm 2$ standard deviation error bars. All four estimators agree with the reference at equilibrium, but Symmetrized VAMPnet is biased for many short off-equilibrium trajectories.
  }
\label{fig:method_compare_pops_timescale}
\end{figure*}

\begin{figure*}[htbp]
\centering
\includegraphics[width=\linewidth]{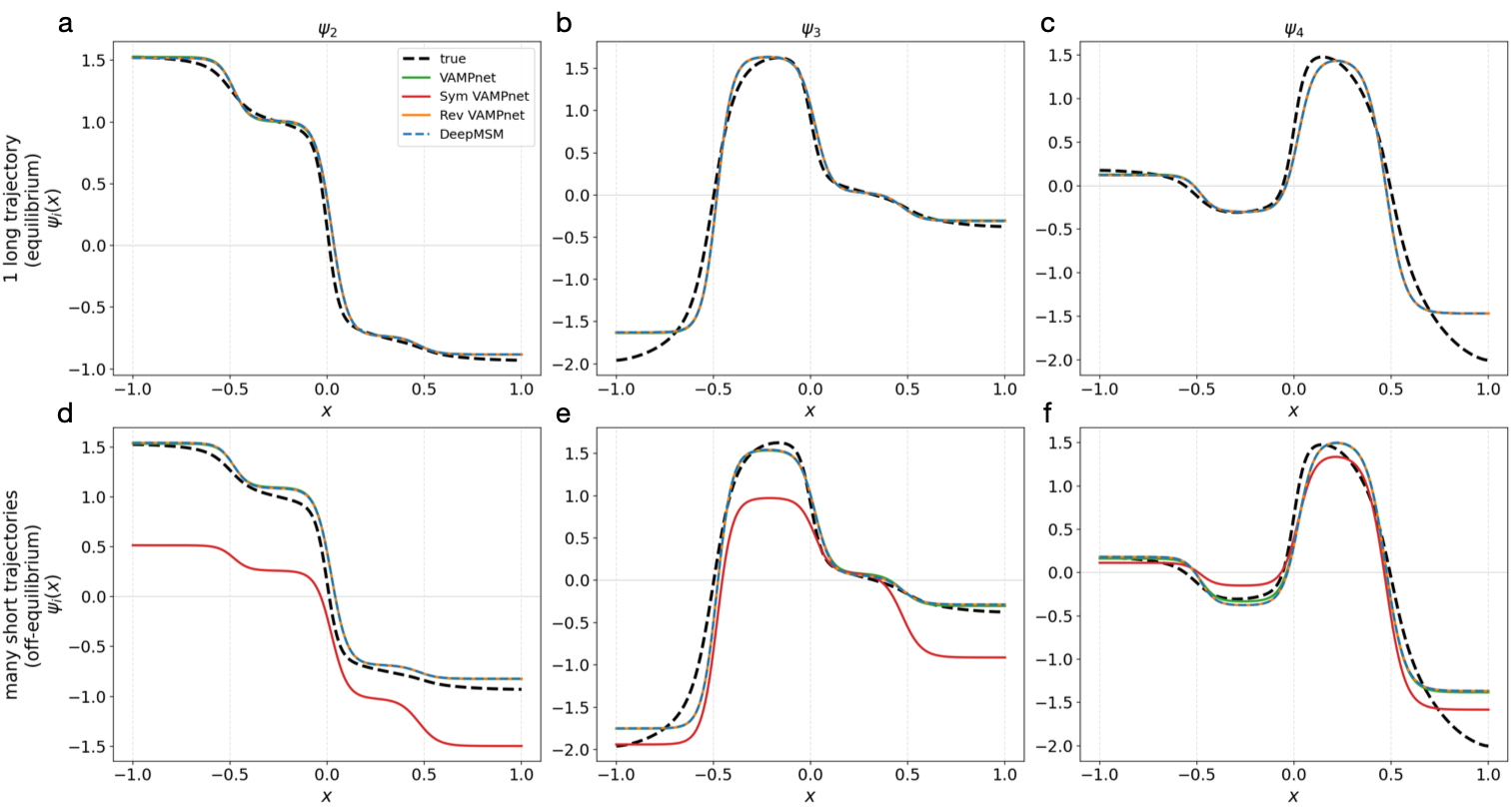}
  \caption{\small The three slowest eigenfunctions of the quadruple-well potential at lag $\tau=10$ frames. (a--c) One long equilibrium trajectory of $10^5$ frames. (d--f) $2\times10^4$ short off-equilibrium trajectories. Dashed black curves are the ground truth. Every estimator recovers the reference from equilibrium data. Only Symmetrized VAMPnet distorts the slow modes for the off-equilibrium data, most visibly in $\psi_2$. DeepMSM (dashed blue) lies over Reversible VAMPnet because both use the same Koopman-reweighted reversible covariances.
  }
\label{fig:method_compare_eigenfunctions}
\end{figure*}

\begin{figure*}[htbp]
\centering
\includegraphics[width=\linewidth]{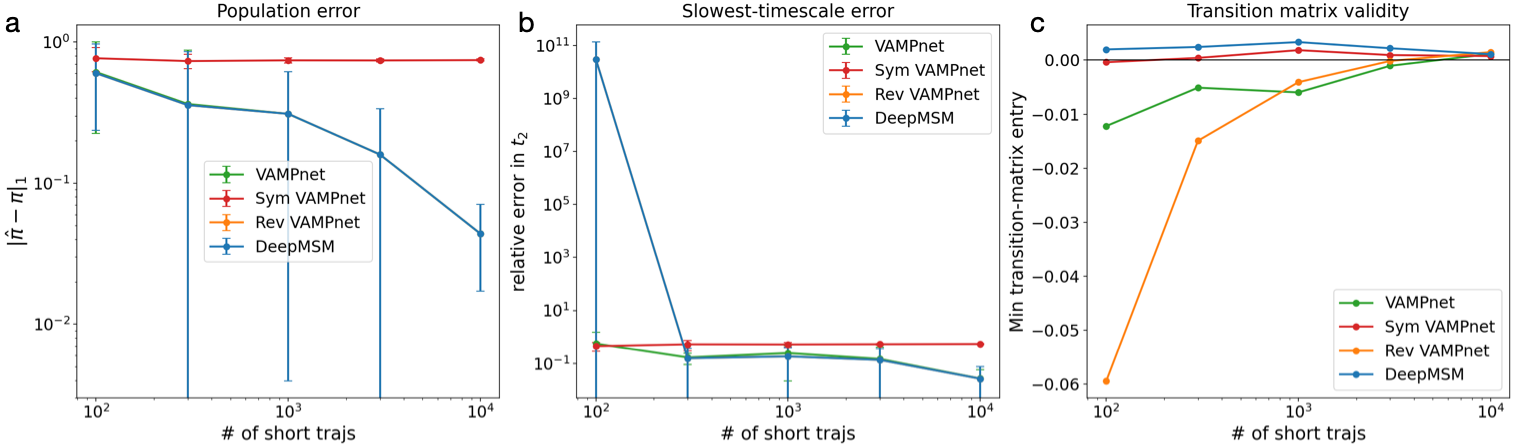}
  \caption{\small Convergence and validity against the number of short off-equilibrium trajectories at $\tau=10$ frames. (a) $\ell_1$ error in the state populations. (b) Relative error in the slowest implied timescale. (c) Minimum entry of the estimated transition matrix, where values below the black line mark a matrix that is not row stochastic. Points are means over five independent trials with $\pm 2$ standard deviation error bars. Symmetrized VAMPnet sits on a bias floor that more data cannot remove, while the other three converge. Only DeepMSM stays nonnegative at every sample size, so it alone is both unbiased off equilibrium and usable as a valid transition matrix. The DeepMSM spike in (b) at $N=10^2$ comes from one divergent independent trial (Fig.~\ref{fig:deepmsm_crispness}a).
  }
\label{fig:method_compare}
\end{figure*}

\begin{figure*}[htbp]
\centering
\includegraphics[width=\linewidth]{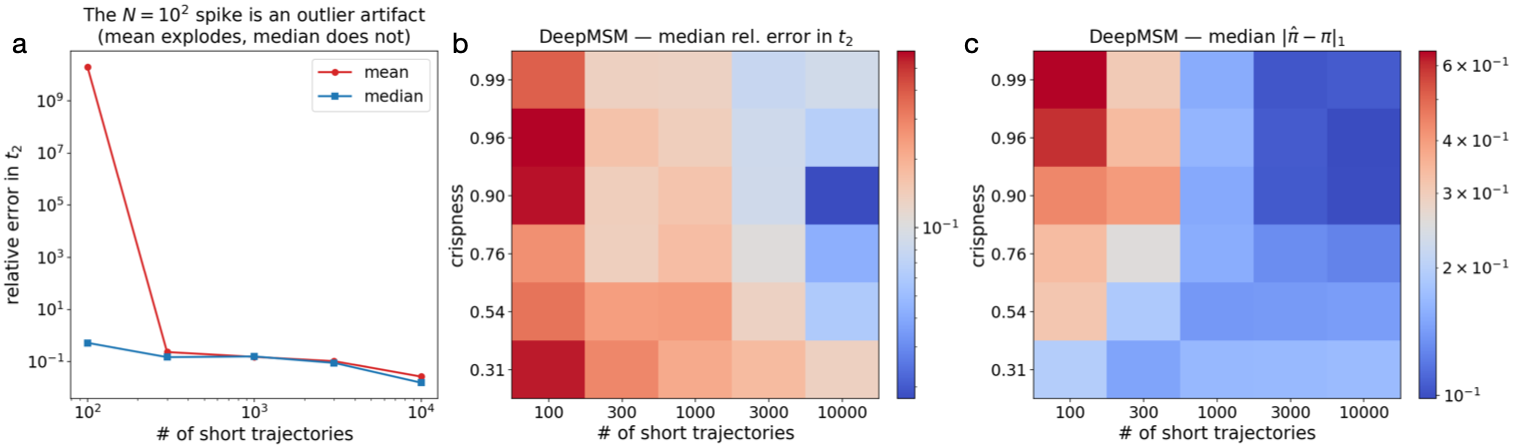}
  \caption{\small The two main breakdown modes of the DeepMSM, namely too few short trajectories and insufficient crispness ($\kappa$) of the state memberships. (a) Relative error in $t_2$ at fixed $\kappa=0.88$, shown as both the mean and the median over independent trials. The two agree except at $N=10^2$, where $\mathbf{C}_{00}$ becomes ill-conditioned, the Koopman weights collapse onto a few frames, and the slow eigenvalue can reach unity so that $t_2=-\tau/\ln\lambda_2$ diverges in independent trials. All later statistics are therefore medians. (b,c) Median error in $t_2$ and in the state populations over a grid of crispness $\kappa$ and trajectory count $N$. Both panels use one fixed reference, the true timescale in (b) and the hard-partition limit of the membership family in (c), so the rows are comparable. Scarce data degrades both observables and is curable by more sampling. Low crispness instead leaves a floor in the populations that sampling cannot remove, while the timescales stay far less sensitive to $\kappa$ because the eigenvalues of $K$ depend on the membership subspace rather than on the individual membership functions.
  }
\label{fig:deepmsm_crispness}
\end{figure*}


\begin{figure*}[htbp]
\centering
\includegraphics[width=\linewidth]{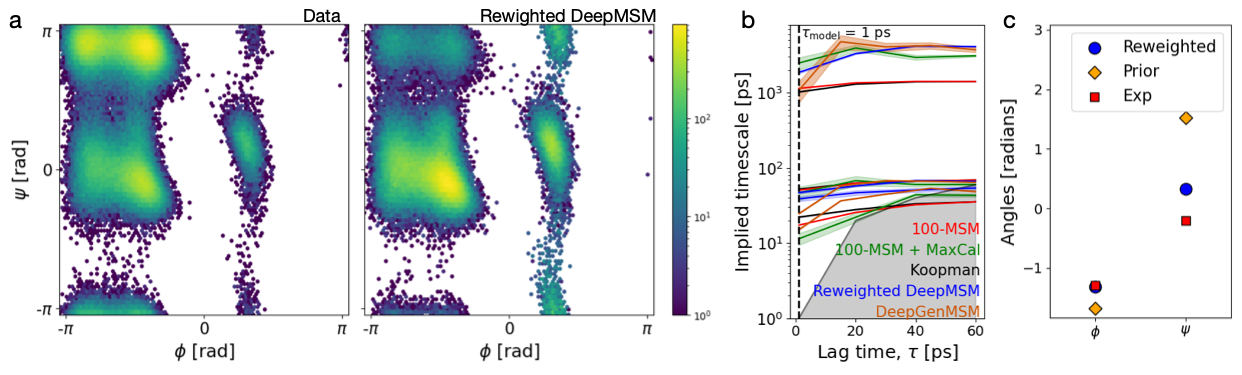}
  \caption{\small   Performance of the BICePs reweighted Deep Reversible MSM in estimating the thermodynamics and kinetics of alanine dipeptide when using dihedral angles ($\phi=-74^\circ$, $\psi=-12^\circ$) as the two experimental observables. (a) Training data distribution and reweighted stationary distribution. (b) Relaxation timescales from the models built using only training data (Koopman and the 100-state MSM), the 100-state MSM + MaxCal reweighting, the reweighted DeepMSM, and the DeepGenMSM. (c) Comparison of ensemble-averaged observables from the prior, BICePs-reweighted, and experimental data.
  }
\label{fig:reweighted_dynamics_angles}
\end{figure*}

\begin{figure*}[htbp]
\centering
\includegraphics[width=\linewidth]{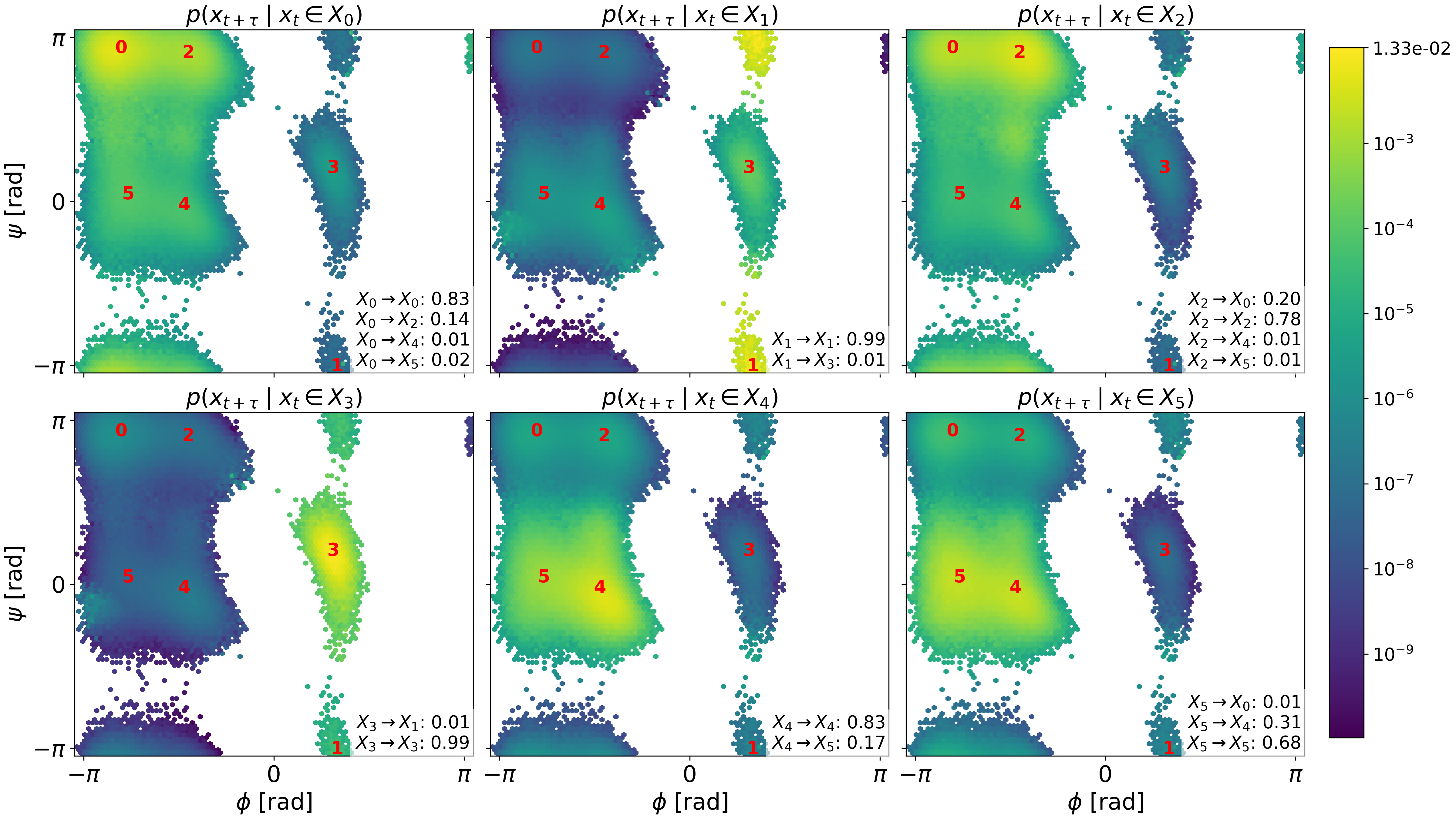}
  \caption{\small  Configurational landing densities, $p(x_{t+\tau} | x_{t} \in X_{i})$  conditioned on BICePs-reweighted populations obtained by applying the $\psi$ and $\psi$ angles as data restraints. The experimental observables were set to be $(\phi=-74^\circ, \psi=-12^\circ)$. State-to-state transition probabilities are reported in the lower-right corner of each panel; values with $K_{ij} < 0.005$ were omitted for clarity.
  }
\label{fig:landing_densities_angles}
\end{figure*}

\begin{figure*}[htbp]
\centering
\includegraphics[width=\linewidth]{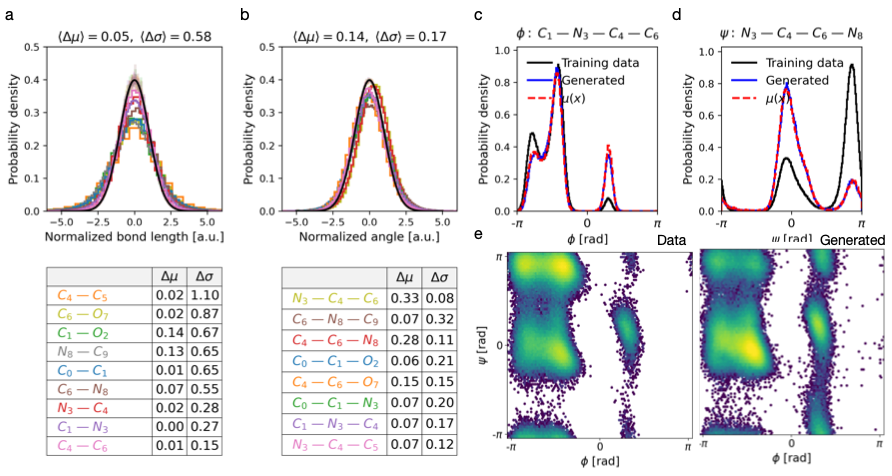}
  \caption{\small Performance of BICePs-reweighted Deep Generative MSM in estimating the thermodynamics and kinetics of alanine dipeptide when given two experimental observables that control the angles ($\phi=-74^\circ$, $\psi=-12^\circ$). (a-b) Normalized bond-length and angle distributions from training data (black curve) and from the trajectory generated from the model. Tables report the differences in the mean, $\Delta \mu$ and standard deviation, $\Delta \sigma$ between the training data and generated trajectory. (c-e) Stationary distributions of $\phi$ and $\psi$. The stationary distribution $\mu$ (red curve) is overlaid to show agreement with the density from the generated trajectory.
  }
\label{fig:gen_phi_psi_angles}
\end{figure*}

\begin{figure*}[htbp]
\centering
\includegraphics[width=\linewidth]{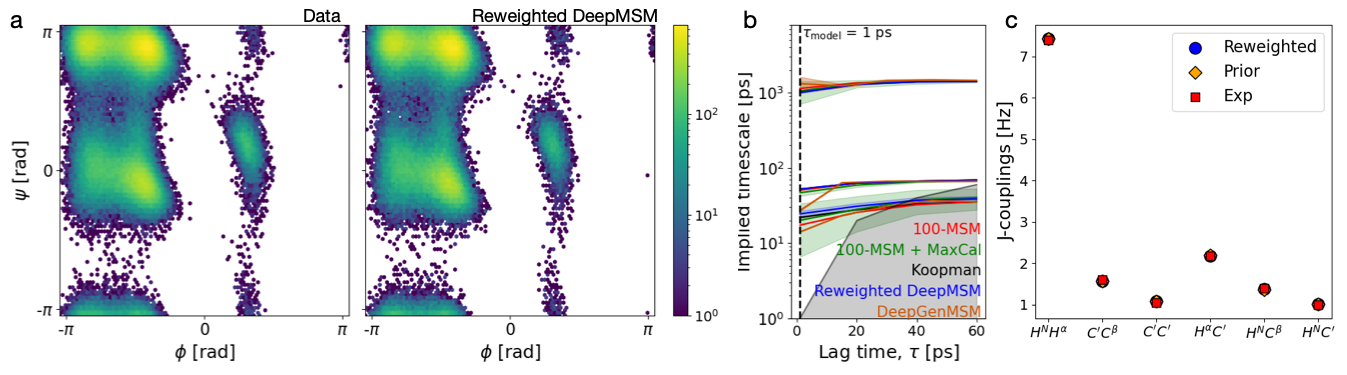}
  \caption{\small Performance of the BICePs reweighted Deep Reversible MSM in estimating the thermodynamics and kinetics of alanine dipeptide when given six experimental J-couplings that are averages over the training data. (a) Training data distribution and reweighted stationary distribution.  (b) Relaxation timescales from the models built using only training data (Koopman and the 100-state MSM), the 100-state MSM + MaxCal reweighting, the reweighted DeepMSM, and the DeepGenMSM. (c) Comparison of ensemble-averaged observables from the prior, BICePs-reweighted, and experimental data.
  }
\label{fig:reweighted_dynamics_J_exp_avg}
\end{figure*}

\begin{figure*}[htbp]
\centering
\includegraphics[width=\linewidth]{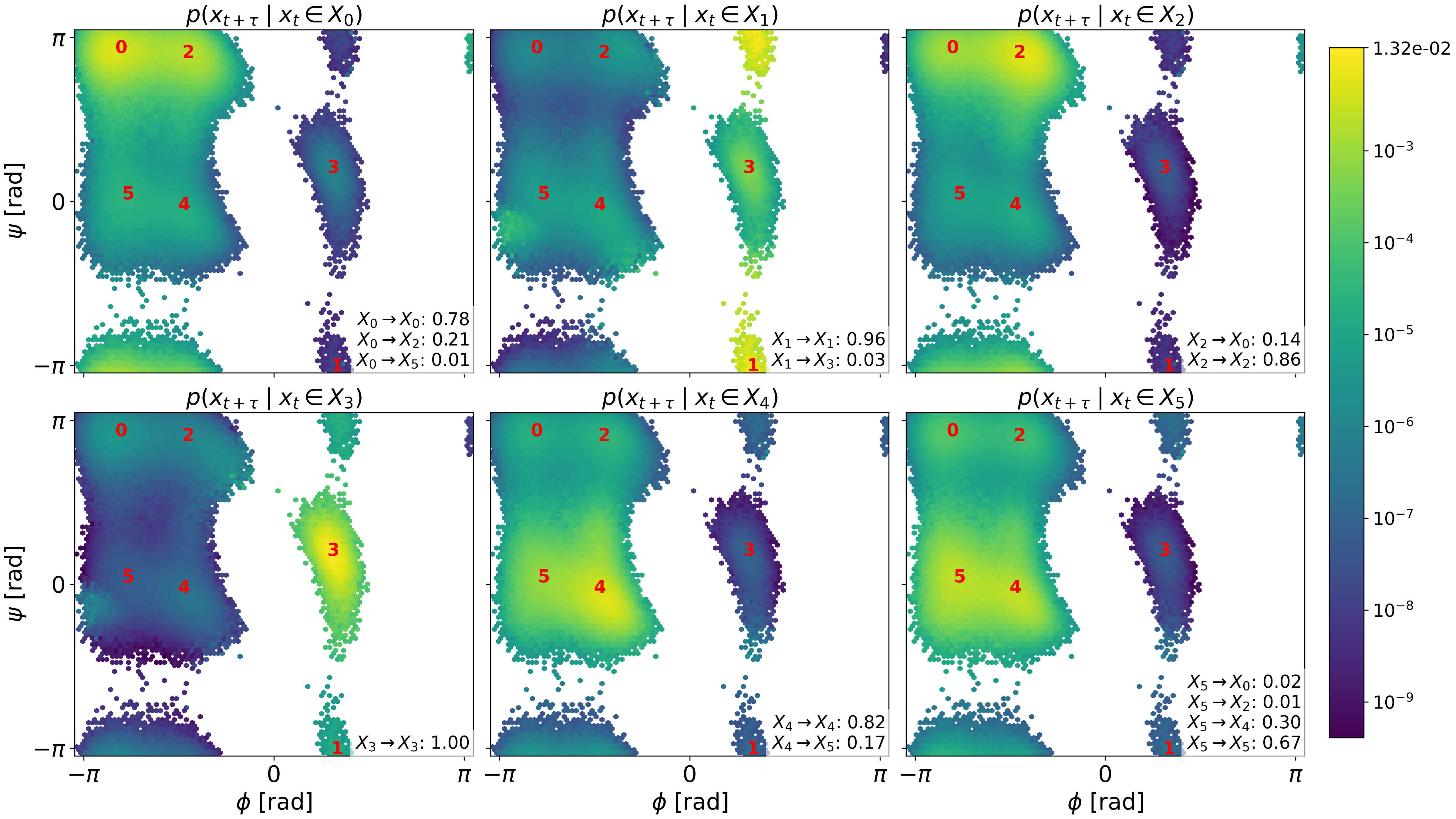}
  \caption{\small  Configurational landing densities, $p(x_{t+\tau} | x_{t} \in X_{i})$  conditioned on BICePs-reweighted populations obtained by applying six J-coupling data restraints. The experimental observables were set equal to the ensemble-averaged J-couplings computed from the alanine dipeptide training data. State-to-state transition probabilities are reported in the lower-right corner of each panel; values with $K_{ij} < 0.005$ were omitted for clarity.
  }
\label{fig:landing_densities_avg}
\end{figure*}

\begin{figure*}[htbp]
\centering
\includegraphics[width=\linewidth]{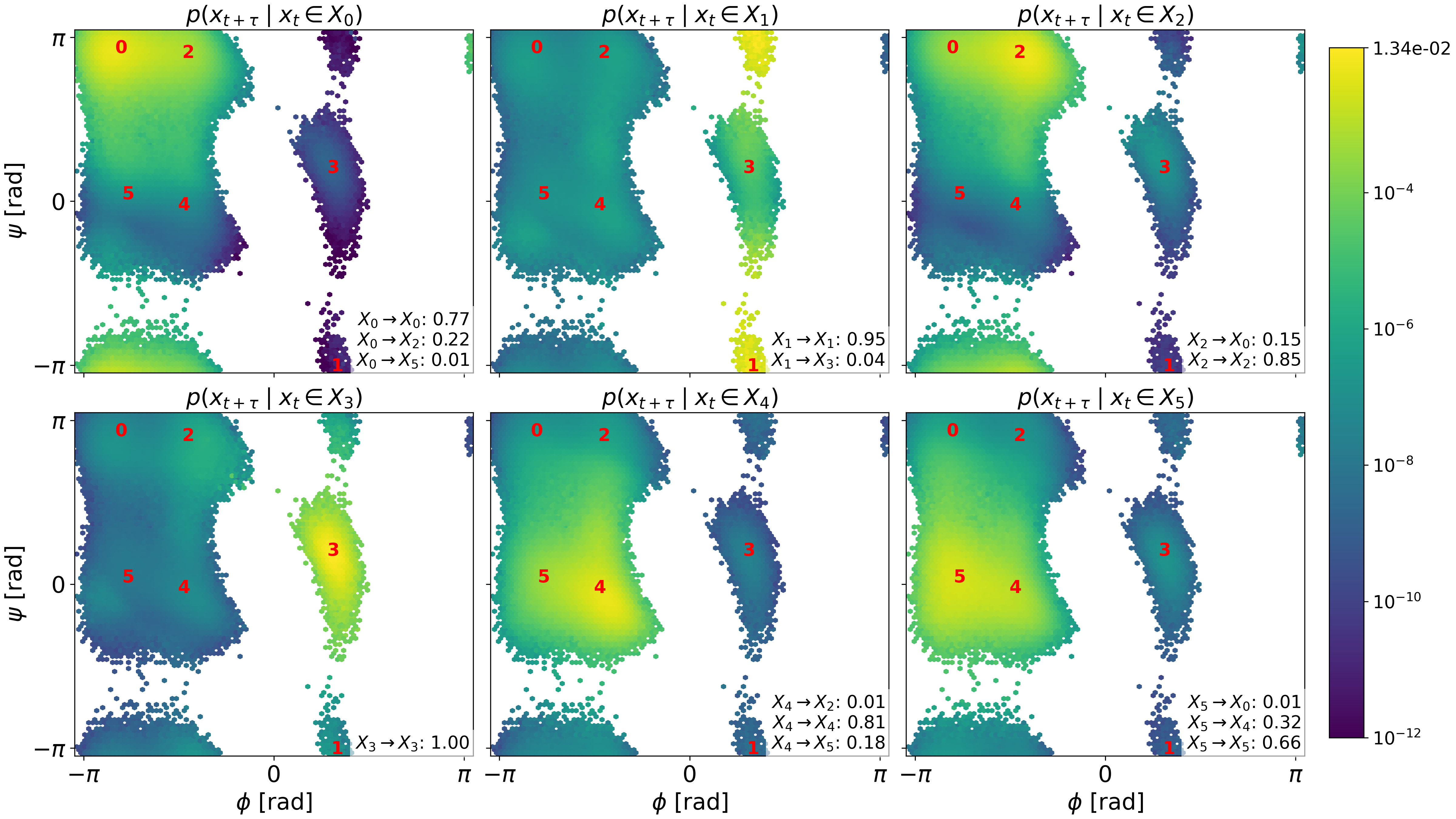}
  \caption{\small   Configurational landing densities of a DeepMSM trained on only trajectory data while using the negative log-likelihood loss from Mardt \textit{et al.}\cite{mardt2020deep}. State-to-state transition probabilities are reported in the lower-right corner of each panel; values with $K_{ij} < 0.005$ were omitted for clarity.
  }
\label{fig:landing_densities_avg_deepMSM_prior}
\end{figure*}

%

\begin{figure*}[htbp]
\centering
\includegraphics[width=\linewidth]{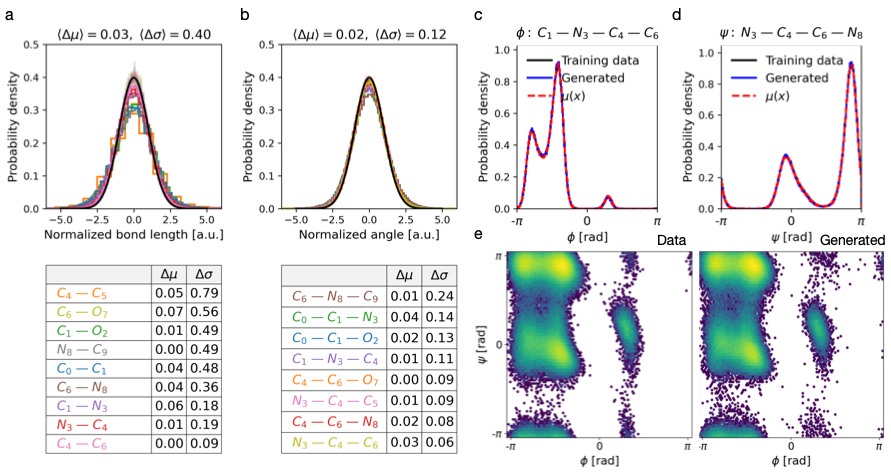}
  \caption{\small Performance of BICePs-reweighted Deep Generative MSM in estimating the thermodynamics and kinetics of alanine dipeptide when given six experimental J-couplings that are averages over the training data.
  (a-b) Normalized bond-length and angle distributions from training data (black curve) and from the trajectory generated from the model. Tables report the differences in the mean, $\Delta \mu$ and standard deviation, $\Delta \sigma$ between the training data and generated trajectory. (c-e) Stationary distributions of $\phi$ and $\psi$. The stationary distribution $\mu$ (red curve) is overlaid to show agreement with the density from the generated trajectory.
  }
\label{fig:gen_J_exp_avg}
\end{figure*}

\begin{figure*}[htbp]
\centering
\includegraphics[width=\linewidth]{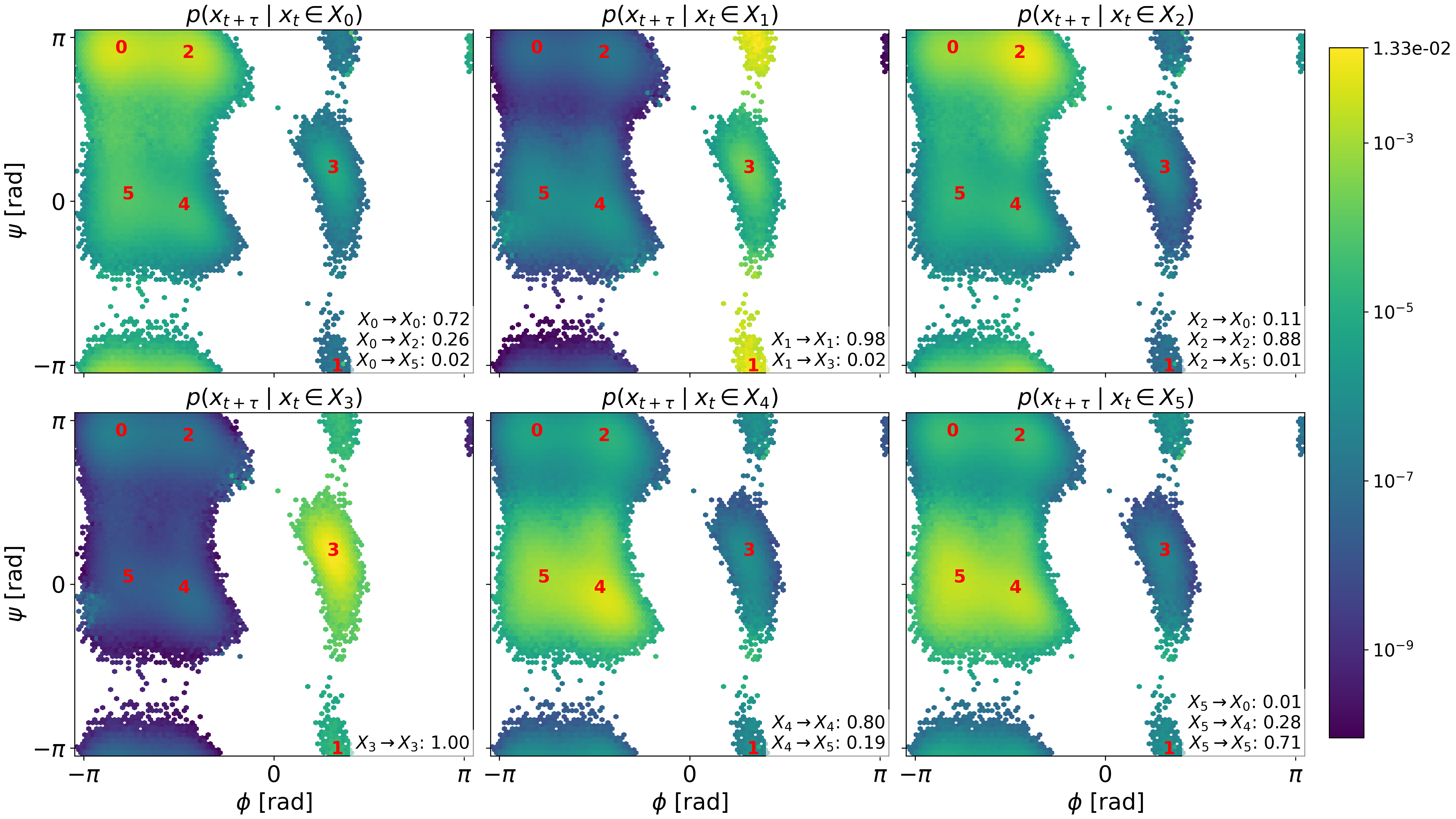}
  \caption{\small  Configurational landing densities, $p(x_{t+\tau} | x_{t} \in X_{i})$  conditioned on BICePs-reweighted populations obtained by applying six J-coupling data restraints. The experimental observables were set equal to J-coupling values computed using $\phi=+60^{\circ}$. State-to-state transition probabilities are reported in the lower-right corner of each panel; values with $K_{ij} < 0.005$ were omitted for clarity.
  }
\label{fig:landing_densities_J_60_deg}
\end{figure*}

\begin{figure*}[htbp]
\centering
\includegraphics[width=\linewidth]{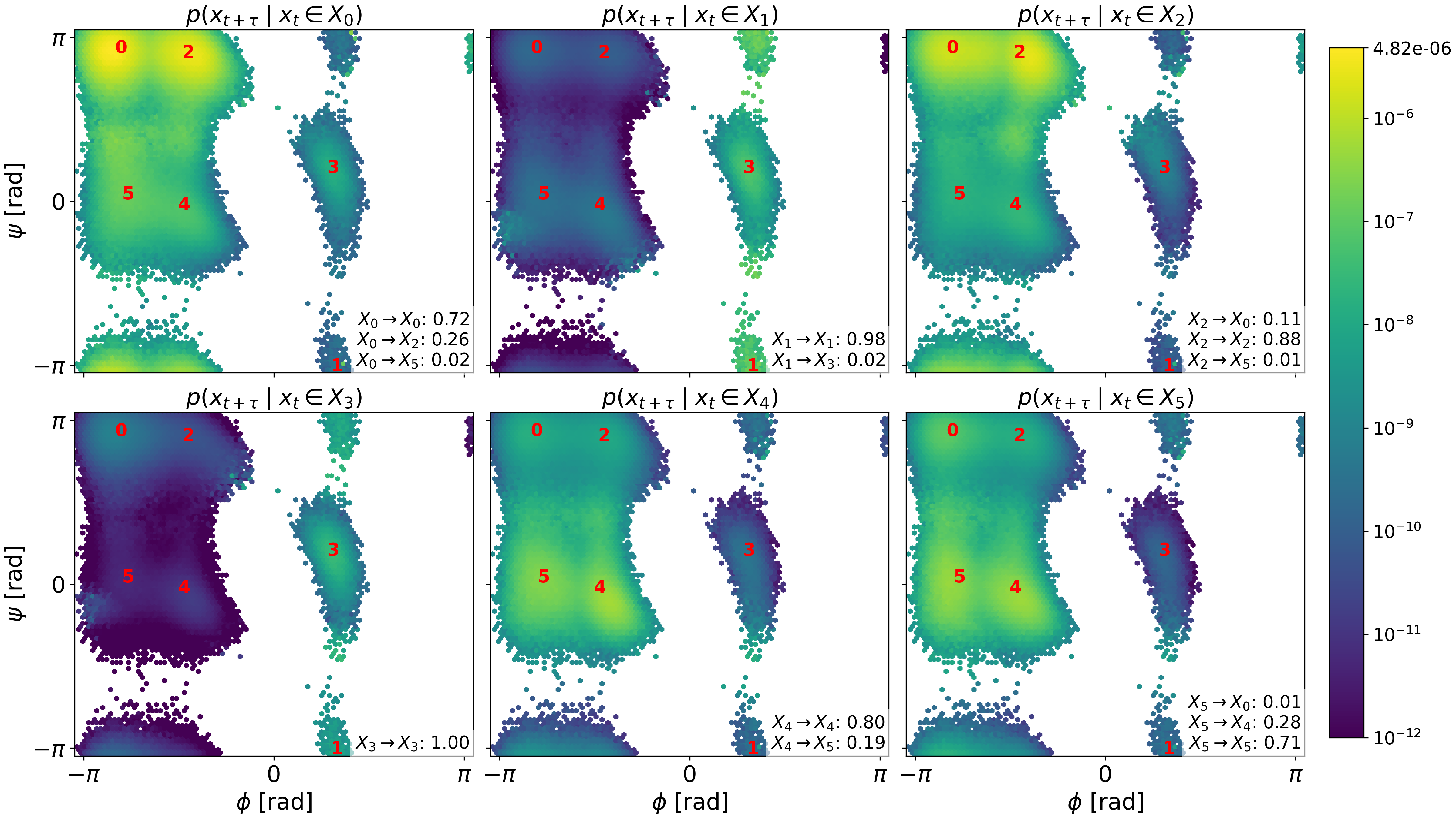}
  \caption{\small  Uncertainty in configurational landing densities when applying six J-coupling data restraints. The experimental observables were set equal to J-coupling values computed using $\phi=+60^{\circ}$.
  }
\label{fig:uncertainty_landing_densities_J_60_deg}
\end{figure*}

\begin{figure*}[htbp]
\centering
\includegraphics[width=\linewidth]{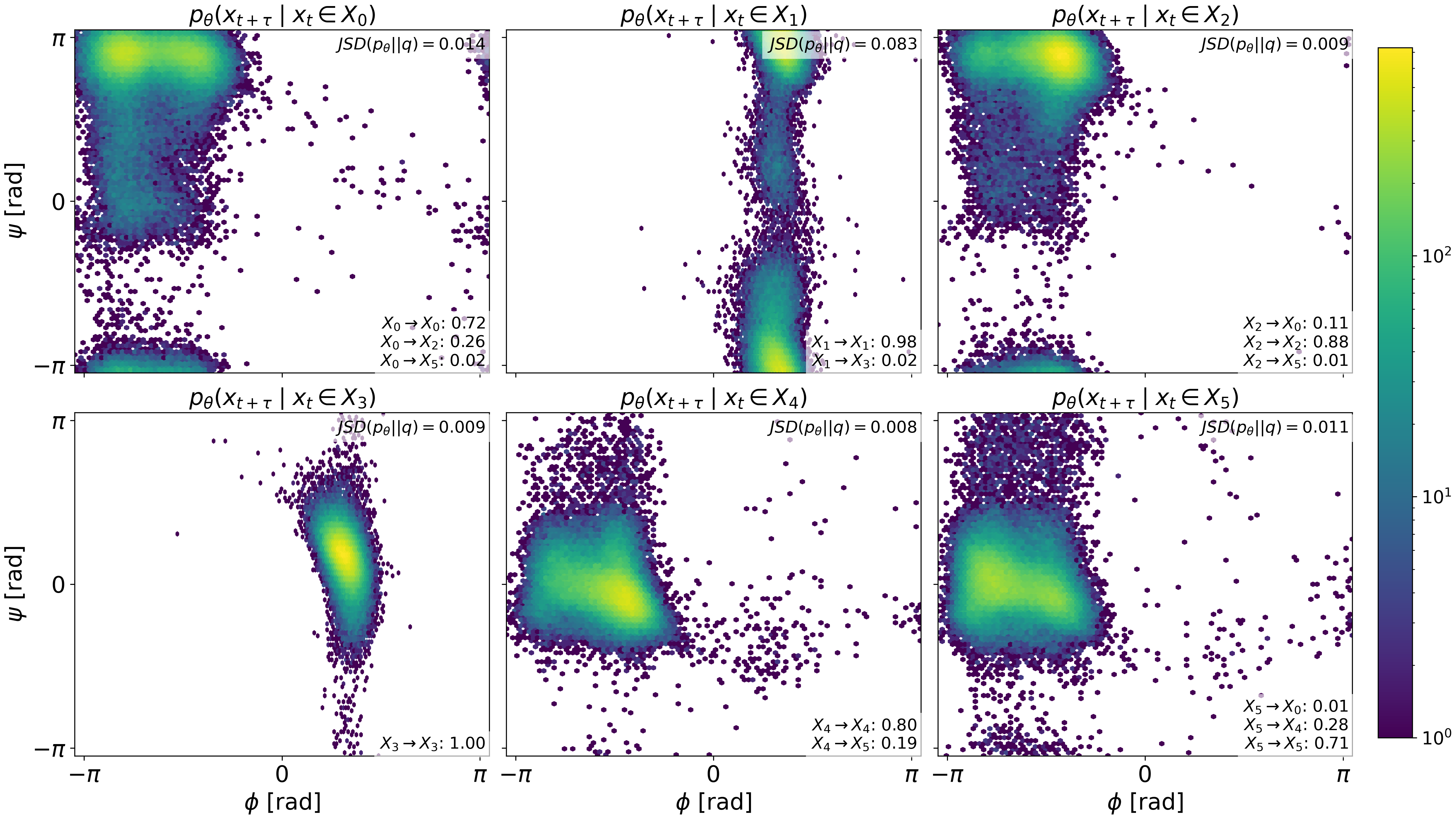}
  \caption{\small  Configurational landing densities from the generative model, $p_{\theta}(x_{t+\tau} | x_{t} \in X_{i})$  trained to mimic the BICePs reweighted DeepMSM landing densities when given six experimental J-coupling data restraints that were set equal to J-coupling values computed at $\phi=+60^{\circ}$. State-to-state transition probabilities are reported in the lower-right corner of each panel; values with $K_{ij} < 0.005$ were omitted for clarity. Jensen--Shannon divergence (JSD) values comparing the generative model landing densities, $p_{\theta}$ to the reweighted DeepMSM  landing densities, $q$ are presented in the top-right of each panel.
  }
  \label{fig:gen_landing_densities_J_60_deg}
\end{figure*}

\end{document}